\documentclass[journal,hidelinks]{IEEEtran}

\usepackage{url}
\usepackage[english]{babel}
\addto\captionsenglish{}
\usepackage[utf8]{inputenc}
\usepackage{amsmath}
\usepackage{amssymb}
\usepackage{mathtools}
\usepackage{cuted}
\usepackage{cite}
\usepackage{graphicx}
\usepackage{xcolor}
\usepackage{hyperref}
\usepackage{tikz}
\usepackage{pgfplots}
\pgfplotsset{compat=newest}
\usepackage{scalefnt}
\usepackage{orcidlink}
\usepackage[acronym,shortcuts]{glossaries}
\usepackage{algorithm, algpseudocode}
\usepackage{algcompatible}
\usepackage{bm}
\usepackage{makecell}
\usepackage{tcolorbox}
\usepackage{multirow}
\usepackage{flushend}
\usepackage[caption=false,font=footnotesize]{subfig}
\usepackage{textcomp}
\usepackage{booktabs}

\newacronym{RMSE}{RMSE}{Root Mean Square Error}
\newacronym{MMSE}{MMSE}{Minimum Mean Square Error}
\newacronym{MF}{MF}{Matched Filter}
\newacronym{RPE}{RPE}{Radar Parameter Estimation}
\newacronym{OTFS}{OTFS}{Orthogonal Time Frequency Space}
\newacronym{AFDM}{AFDM}{Affine Frequency Division Multiplexing}
\newacronym{MIMO}{MIMO}{Multiple-Input Multiple-Output}
\newacronym{SISO}{SISO}{Single-Input Single-Output}
\newacronym{ISAC}{ISAC}{Integrated Sensing and Communications}
\newacronym{3D}{3D}{Three-Dimensional}
\newacronym{2D}{2D}{Two-Dimensional}
\newacronym{1D}{1D}{One-Dimensional}
\newacronym{RX}{RX}{Receiver}
\newacronym{TX}{TX}{Transmitter}
\newacronym{BF}{BF}{Beamforming}
\newacronym{mmWave}{mmWave}{Millimeter-Wave}
\newacronym{SotA}{SotA}{State-of-the-Art}
\newacronym{ULA}{ULA}{Uniform Linear Array}
\newacronym{QAM}{QAM}{Quadrature Amplitude Modulation}
\newacronym{ISFFT}{ISFFT}{Inverse Symplectic Finite Fourier Transform}
\newacronym{SFFT}{SFFT}{Symplectic Finite Fourier Transform}
\newacronym{AWGN}{AWGN}{Additive White Gaussian Noise}
\newacronym{OFDM}{OFDM}{Orthogonal Frequency Division Multiplexing}
\newacronym{OCDM}{OCDM}{Orthogonal Chirp Division Multiplexing}
\newacronym{BS}{BS}{Base Station}
\newacronym{UE}{UE}{User Equipment}
\newacronym{DFT}{DFT}{Discrete Fourier Transform}
\newacronym{IDFT}{IDFT}{Inverse Discrete Fourier Transform}
\newacronym{IFFT}{IFFT}{Inverse Fast Fourier Transform}
\newacronym{TD}{TD}{Time-Domain}
\newacronym{wlg}{wlg}{Without Loss of Generality}
\newacronym{CP}{CP}{Cyclic Prefix}
\newacronym{DAFT}{DAFT}{Discrete Affine Fourier Transform}
\newacronym{DAF}{DAF}{Discrete Affine Fourier}
\newacronym{IDAFT}{IDAFT}{Inverse Discrete Affine Fourier Transform}
\newacronym{CPP}{CPP}{\textit{Chirp-Periodic} Prefix}
\newacronym{IDZT}{IDZT}{Inverse Discrete Zak Transform}
\newacronym{DZT}{DZT}{Discrete Zak Transform}
\newacronym{ICI}{ICI}{Inter-Carrier Interference}
\newacronym{BER}{BER}{Bit Error Rate}
\newacronym{DoF}{DoF}{Degrees-of-Freedom}
\newacronym{FD}{FD}{Full-Duplex}
\newacronym{SIMO}{SIMO}{Single-Input Multiple-Output}
\newacronym{MISO}{MISO}{Multiple-Input Single-Output}
\newacronym{AoD}{AoD}{Angle-of-Departure}
\newacronym{AoA}{AoA}{Angle-of-Arrival}
\newacronym{RF}{RF}{Radio Frequency}
\newacronym{SIM}{SIM}{Stacked Intelligent Metasurfaces}
\newacronym{FPGA}{FPGA}{Field Programmable Gate Array}
\newacronym{UPA}{UPA}{Uniform Planar Array}
\newacronym{CC}{CC}{Communication-Centric}
\newacronym{I/O}{I/O}{Input-Output}
\newacronym{iid}{i.i.d.}{Independent and Identically Distributed}
\newacronym{IoT}{IoT}{Internet of Things}
\newacronym{V2X}{V2X}{Vehicle-to-Everything}
\newacronym{NTN}{NTN}{Non-Terrestrial Network}
\newacronym{LEO}{LEO}{Low-Earth Orbit}
\newacronym{THz}{THz}{Terahertz}
\newacronym{EM}{EM}{Expectation Maximization}
\newacronym{RIS}{RIS}{Reconfigurable Intelligent Surface}
\newacronym{DoA}{DoA}{Direction-of-Arrival}
\newacronym{DD}{DD}{Doubly-Dispersive}
\newacronym{ODDM}{ODDM}{Orthogonal Delay-Doppler Division Multiplexing}
\newacronym{LoS}{LoS}{Line-of-Sight}
\newacronym{NLoS}{NLoS}{Non-Line-of-Sight}
\newacronym{6G}{6G}{Sixth Generation}
\newacronym{MPDD}{MPDD}{Metasurfaces-Parametrized DD}
\newacronym{GaBP}{GaBP}{Gaussian Belief Propagation}
\newacronym{MSE}{MSE}{Mean-Squared-Error}
\newacronym{sIC}{soft IC}{Soft Interference Cancellation}
\newacronym{soft RG}{soft RG}{Soft Replica Generation}
\newacronym{BG}{BG}{Belief Generation}
\newacronym{SGA}{SGA}{Scalar Gaussian Approximation}
\newacronym{CLT}{CLT}{Central Limit Theorem}
\newacronym{PDF}{PDF}{Probability Density Function}
\newacronym{QPSK}{QPSK}{Quadrature Phase-Shift Keying}
\newacronym{OQAM}{OQAM}{Offset Quadrature Amplitude Modulation}
\newacronym{LMMSE}{LMMSE}{Linear Minimum Mean Square Error}
\newacronym{SNR}{SNR}{Signal-to-Noise Ratio}
\newacronym{OOBE}{OOBE}{Out-of-Band Emission}
\newacronym{PAPR}{PAPR}{Peak-to-Average Power Ratio}
\newacronym{AFBM}{AFBM}{Affine Filter Bank Modulation}
\newacronym{FBMC}{FBMC}{Filter Bank Multicarrier Modulation}
\newacronym{PPN}{PPN}{Polyphase Network}
\newacronym{SIR}{SIR}{Signal-to-Interference Ratio}
\newacronym{AF}{AF}{Ambiguity Function}
\newacronym{PDA}{PDA}{Probabilistic Data Association}
\newacronym{SBL}{SBL}{Sparse Bayesian Learning}
\newacronym{VGA}{VGA}{Vector Gaussian Approximation}
\newacronym{KL}{KL}{Kullback-Leibler}
\newacronym{GAMP}{GAMP}{Generalized Approximate Message Passing}
\newacronym{EP}{EP}{Expectation Propagation}
\newacronym{5G}{5G}{Fifth Generation}
\newacronym{4G}{4G}{Fourth Generation}
\newacronym{IBO}{IBO}{Input Back-Off}
\newacronym{HPA}{HPA}{High-Power Amplifier}
\newacronym{PA}{PA}{Power Amplifier}
\newacronym{SSPA}{SSPA}{Solid-State Power Amplifier}
\newacronym{PSLR}{PSLR}{Peak Sidelobe Level Ratio}
\newacronym{ISLR}{ISLR}{Integrated Sidelobe Level Ratio}

\newcommand\scalemath[2]{\scalebox{#1}{\mbox{\ensuremath{\displaystyle #2}}}}

\newcommand{\herm}[0]{^{\mathsf{H}}}

\begin{document}

\title{Distortion-Aware Integrated Sensing and Communication with Affine Filter Bank Modulation}


\author{Eya Gourar\textsuperscript{\orcidlink{0009-0001-9575-6075}},
Henrique L. Senger\textsuperscript{\orcidlink{0009-0004-1586-8168}},
Gustavo P. Gonçalves\textsuperscript{\orcidlink{0009-0000-8260-4390}}, \\
Kuranage Roche Rayan Ranasinghe\textsuperscript{\orcidlink{0000-0002-6834-8877}}, \IEEEmembership{Graduate Student Member,~IEEE,}
Hyeon Seok Rou\textsuperscript{\orcidlink{0000-0003-3483-7629}}, \IEEEmembership{Member,~IEEE,} \\
Bruno S. Chang\textsuperscript{\orcidlink{0000-0003-0232-7640}}, 
\IEEEmembership{Member,~IEEE,}
Yahia Medjahdi\textsuperscript{\orcidlink{0000-0001-7030-1130}}, \IEEEmembership{Member,~IEEE,} \\
Giuseppe Thadeu Freitas de Abreu\textsuperscript{\orcidlink{0000-0002-5018-8174}}, \IEEEmembership{Senior Member,~IEEE,} and Didier Le Ruyet\textsuperscript{\orcidlink{0000-0002-9673-2075}}, \IEEEmembership{Senior Member,~IEEE}
\vspace{-3ex}

\thanks{E.~Gourar and Y.~Medjahdi are with  IMT Nord Europe, Institut Mines Télécom, Center for Digital Systems, F-59653 Villeneuve d’Ascq, France (emails:  [eya.gourar,yahia.medjahdi]@imt-nord-europe.fr).} 
\thanks{H.~L.~Senger, G.~P.~Gonçalves and B.~S.~Chang are with the CPGEI/Electronics Department, Federal University of Technology - Paraná, Curitiba, Brazil (emails: [hsenger,gustavog.1999]@alunos.utfpr.edu.br, bschang@utfpr.edu.br).}
\thanks{K.~R.~R.~Ranasinghe, H~S.~Rou and G.~T.~F.~de~Abreu are with the School of Computer Science and Engineering, Constructor University, Campus Ring 1, 28759 Bremen, Germany (emails: [kranasinghe,hrou, gabreu]@constructor.university).}
\thanks{D.~Le~Ruyet is with the CEDRIC, Conservatoire National des Arts et Métiers - Paris, France (email: didier.le$\_$ruyet@cnam.fr).}
\thanks{Parts of this work have been accepted for presentation at the 2026 Asilomar Conference on Signals, Systems, and Computers~\cite{gourar2026robustness}.}
}


\maketitle

\begin{abstract} 
The stringent energy-efficiency requirements of future \ac{ISAC} systems are fundamentally challenged. Unlike conventional communication systems, \ac{ISAC} transmitters must radiate significantly higher power to ensure reliable target detection, forcing the \ac{HPA} to operate closer to saturation, where nonlinear distortions become unavoidable. Consequently, the robustness of every candidate \ac{ISAC} waveform to \ac{HPA} nonlinearities must be carefully assessed. In this context, this paper investigates the robustness of \ac{AFBM}, a recently proposed waveform that combines the delay-Doppler resilience of affine modulation with reduced \ac{PAPR} and improved spectral containment. We develop a statistical characterization of the \ac{AF} of the amplified \ac{AFBM} waveform, deriving approximate expressions for its mean, variance, and Rician-distributed magnitude. Furthermore, a low-complexity Gaussian belief propagation receiver accounting for \ac{HPA} nonlinearities is proposed for communication detection. Simulation results validate the analytical framework and demonstrate that \ac{AFBM} preserves favorable sensing characteristics and robust \ac{BER} performance even under severe nonlinear amplification.
\end{abstract}


\IEEEpeerreviewmaketitle

\glsresetall

\section{Introduction}

\ac{ISAC} aims to use wireless transmission to simultaneously convey information and probe the surrounding environment~\cite{liu2020joint}. By unifying these two functionalities, \ac{ISAC} enables applications such as localization, tracking, and context awareness without dedicating separate spectrum or hardware chains for sensing. 

These new functionalities create waveform requirements that go beyond traditional communication metrics, like throughput and reliability, since the transmitted signal must also exhibit favorable sensing ambiguity properties, robust spectral use, and robustness to practical impairments.

While \ac{OFDM} is expected to remain the multicarrier solution for 6G systems due to its simplicity and mature ecosystem, it is also known to exhibit high \ac{PAPR}, suboptimal performance on doubly-dispersive channels, and non-negligible \ac{OOBE}. These characteristics become particularly problematic in \ac{ISAC} systems, especially in high-mobility scenarios such as Vehicle-to-Vehicle communications. These limitations motivate the study of alternative waveforms that preserve implementation tractability while improving time-frequency localization for both communications and sensing~\cite{rou2024orthogonal}.

In this context, \ac{AFDM} has emerged as a promising alternative modulation scheme by exploiting affine (chirp) structures~\cite{yin2025ofdm}. The use of chirps can be understood from the representation of doubly-dispersive channels. Rather than seeking a fixed orthonormal basis that diagonalizes the channel, the objective is to choose a signal domain in which the channel response becomes structured and sparse. Such a representation is possible because practical propagation channels typically occupy a bounded region in the delay-Doppler plane~\cite{rou2026resurrection}. These features have led to numerous recent developments in \ac{AFDM}-based ISAC, including a thorough analysis of its \ac{AF} and the corresponding transceiver design (see $e.g.$,~\cite{rou2025normalized,ni2025ambiguity,ni2025integrated,luo2025novel}) which support its use as a novel \ac{ISAC} waveform.

Nevertheless, \ac{AFDM} inherits the high PAPR and poor spectral containment of OFDM. To address this gap, a novel waveform, termed \ac{AFBM}, was first introduced in~\cite{senger2025affine}, which combines affine-domain spreading with per-subcarrier filtering to reduce the \ac{OOBE} and the \ac{PAPR} of the \ac{AFDM} signal without excessively introducing transceiver complexity, while keeping its advantages. The current state of the art indicates that AFBM preserves quasi-orthogonality and strong resilience to doubly dispersive channels, while achieving approximately 3~dB lower \ac{PAPR} than conventional AFDM and  \ac{OOBE} as low as -100~dB with PHYDYAS prototype filters~\cite{bellanger2001specification}. Recent studies also report competitive \ac{BER}, \ac{AF}, and sensing-estimation performance using tailored Gaussian belief-propagation detection and probabilistic target-estimation methods~\cite{ranasinghe2025affinefilterbankmodulation}.

Even though \ac{AFBM} improves \ac{PAPR}, its real value depends on how it handles hardware imperfections. In \ac{ISAC} systems, the sensing functionality imposes significantly higher transmit power requirements than conventional communications, since radar echoes undergo two-way propagation losses and additional attenuation from target reflections, resulting in weak received signals~\cite{cui2024energy}. Consequently, \ac{HPA} must operate at high output powers, often close to saturation, to guarantee sufficient sensing coverage and target detectability. However, this operating regime inherently introduces nonlinear distortion, which becomes particularly problematic for the high-PAPR waveforms commonly employed in modern communication systems, as their large envelope fluctuations frequently drive the \ac{HPA} into its nonlinear region~\cite{guel2009etude}.

This raises a central question: does the lower \ac{PAPR} of AFBM translate into increased robustness to \ac{HPA} nonlinearities, or does nonlinear distortion remain a major factor limiting performance? Moreover, although operating the \ac{HPA} with substantial back-off can mitigate these distortions by keeping the amplifier in its linear operating region, it substantially reduces power efficiency, leading to greater energy dissipation and limiting the achievable transmit power. Therefore, \ac{ISAC} systems face an inherent trade-off between transmit power, amplifier linearity, and energy efficiency, making \ac{HPA} nonlinearities an unavoidable hardware impairment whose impact must be carefully analyzed to understand their effect on both communication reliability and sensing performance.

Research on the impact of \ac{HPA} nonlinearities on the AFBM framework remains at an early stage. In our previous work in~\cite{gourar2026robustness}, we investigated the influence of such distortions on the \ac{AF} and on sensing performance when employing the \ac{PDA}-based sensing algorithm proposed in~\cite{ranasinghe2025affinefilterbankmodulation}. Our results demonstrated that both waveform-level and receiver-level sensing performance are largely insensitive to PA-induced distortions, thereby highlighting the robustness and practical advantages of AFBM in realistic \ac{ISAC} system scenarios. However, a comprehensive characterization of the gain associated with the AFBM waveform in the presence of such hardware impairments should also incorporate the communication aspects, which have not yet been investigated. Moreover, an analytical framework capable of describing the typical behavior and fluctuations of the \ac{AF} is needed to obtain a more complete understanding of the AFBM sensing robustness to nonlinear amplification. 

This work presents an analysis of the AFBM waveform under \ac{HPA}-induced distortions. Our main contributions are:
\begin{itemize}
    \item We develop a statistical characterization of the \ac{AF} of the AFBM waveform under \ac{HPA} nonlinearities using the Bussgang decomposition followed by a memoryless polynomial approximation. Approximate expressions for the mean and variance are derived, showing that the AF magnitude remains Rician distributed after nonlinear amplification. The analysis is validated through simulations.
    \item The communication performance robustness of \ac{AFBM} under \ac{HPA} nonlinearities is investigated. Numerical results demonstrate that AFBM maintains robust \ac{BER} performance even under severe nonlinear amplification, showcasing the resilience of the novel waveform.
    \item A low-complexity \ac{HPA}-aware Gaussian belief propagation receiver is derived by explicitly accounting for the nonlinear distortion. The proposed receiver mitigates the impact of \ac{HPA} nonlinearities and provides improved communication performance with only a marginal increase in computational complexity.
\end{itemize}

The remainder of this paper is organized as follows: Section~\ref{sec:system_model} introduces the system model and the \ac{AFBM} transmit signal construction, while Section~\ref{sec:af_analysis} derives the ambiguity function considering power amplifier nonlinearities. Section~\ref{sec:receivers} presents the communications receiver algorithms for the proposed architecture. The simulation results are presented and analyzed in Section~\ref{sec:results}. Finally, the concluding remarks are given in Section~\ref{sec:conclusion}.

\textit{Notations:}~ The following notations are adopted throughout this manuscript. Boldface lowercase and uppercase letters denote vectors and matrices, respectively. The sets of $N$-dimensional complex and real column vectors are denoted by $\mathbb{C}^{N}$ and $\mathbb{R}^{N}$, respectively, while $\mathbb{C}^{M\times N}$ and $\mathbb{R}^{M\times N}$ denote the corresponding sets of $M\times N$ matrices. For a vector $\mathbf{a}$, the notation $a(n)$ denotes its $n$-th element, while for a matrix $\mathbf{A}$, $A(m,n)$ denotes the entry in its $m$-th row and $n$-th column. The operators $(\cdot)^*$, $(\cdot)^T$, and $(\cdot)^H$ denote complex conjugation, transpose, and Hermitian (conjugate transpose), respectively. For a given matrix $\bar{\mathbf{A}}$, $\bar{\mathbf{A}}^{\dagger}$ denotes its left Moore--Penrose pseudoinverse. The notation $\mathcal{CN}(\mu,\sigma^2)$ denotes the circularly symmetric complex Gaussian distribution with mean $\mu$ and variance $\sigma^2$. If $Z\sim\mathcal{CN}(\mu,\sigma^2)$, then $|Z|\sim\mathrm{Rice}(|\mu|,\sigma^2/2)$. The identity matrix of size $N$ is denoted by $\mathbf{I}_N$, the all-zero vector by $\mathbf{0}_N$ and $\mathbf{0}_{L \times N}$ denotes an all-zero matrix of size $L\times N$. $\mathbf{F}_{L}$ denotes the normalized $L$-point \ac{DFT} matrix. The operator $\Re(\cdot)$ extracts the real part of a complex quantity. $\otimes$ and $\odot$ denote the Kronecker product and the Hadamard (element-wise) product, respectively. For a matrix $\mathbf{A}$, $|\mathbf{A}|$ denotes the element-wise magnitude, while $|\mathbf{A}|^{\odot p}$ denotes the element-wise $p$-th power of the magnitude, $i.e.$, $\left[|\mathbf{A}|^{\odot p}\right]_{m,n}=|A(m,n)|^p$. The expectation operator is denoted by $\mathbb{E}[\cdot]$. The operators $\operatorname{diag}(\mathbf{A})$ and $\operatorname{diag}(\mathbf{a})$ denote, respectively, the vector formed by the diagonal entries of a matrix and the diagonal matrix whose diagonal entries are given by the vector $\mathbf{a}$. The Euclidean norm of a vector is denoted by $\|\cdot\|_2$, while the Frobenius norm of a matrix is denoted by $\|\cdot\|_F$, and $\operatorname{tr}(\cdot)$ denotes the trace operator.

\section{System Model} \label{sec:system_model}


\begin{figure*}[t!]
    \centering
    \makebox[\textwidth][c]{%
        \resizebox{1.0\textwidth}{!}{%
%
\usetikzlibrary{arrows.meta}
\definecolor{cGray}{HTML}{CCCCCC}   
\definecolor{cYellow}{HTML}{FFEEAA} 
\definecolor{cGreen}{HTML}{AAFFAA}  
\definecolor{cOrange}{HTML}{FF9955} 
\definecolor{cCyan}{HTML}{AAFFEE}   
\definecolor{cLime}{HTML}{CCFFAA}   
\definecolor{cChan}{HTML}{E6E6E6}   
\definecolor{cLabel}{HTML}{5A6E5A}  

%
\newcommand{\flagblock}[8][0pt]{%
    \pgfmathsetmacro{\bh}{1.5}%
    \pgfmathsetmacro{\lft}{#3-#5/2}%
    \pgfmathsetmacro{\rgt}{#3+#5/2}%
    \pgfmathsetmacro{\yt}{#4+\bh/2}%
    \pgfmathsetmacro{\yb}{#4-\bh/2}%

    \node (#2) at (#3,#4) {};%

    \draw[
        fill=#6,
        draw=black,
        line width=0.5pt,
        rounded corners=5pt
    ]
    (\lft,\yb) rectangle (\rgt,\yt);

    \node[
        align=center,
        anchor=south,
        font=\footnotesize\bfseries,
        yshift=#1
    ] at (#3,#4+0.03) {#7};

    \node[
        align=center,
        anchor=north,
        font=\footnotesize
    ] at (#3,#4-0.10) {#8};
}

\newcommand{\dottedbox}[4]{%
    \draw[->][
        black,
        densely dotted,
        line width=0.7pt
    ]
    (#1,#2) rectangle (#3,#4);
}

\newcommand{\boxtitle}[3]{%
    \node[
        anchor=south west,
        font=\small\itshape,
        color=cLabel,
        fill=white,
        inner sep=1pt
    ] at (#1,#2) {#3};
}

\begin{tikzpicture}[>=latex,line width=0.6pt]

\begin{scope}[shift={(0.98,0)}]


\dottedbox{0.15}{0.95}{15.55}{-0.95}

\dottedbox{5.65}{-2.55}{12.65}{-4.45}

\draw[
    fill=cChan,
    draw=black,
    line width=0.5pt,
    rounded corners=2pt
]
(16.15,1.05) rectangle (17.75,-4.35);

\node[
    font=\bfseries\Large
] at (16.95,-3.0)
{$\mathbf{H}$};

\node[
    rotate=90,
    font=\small,
    align=center
] at (16.95,-0.65)
{Doubly-dispersive\\Channel};


\flagblock{mapping}
{1.55}{0}{2.4}
{cGray}
{Mapping}
{$\bm{\Xi}$}

\flagblock{precod}
{4.55}{0}{2.8}
{cYellow}
{DAFT Precoding}
{$\mathbf{I}_{\mathbf K}
 \otimes
 \mathbf{C}_\mathbf{f}$}

\flagblock[-1.2mm]{idaft}
{7.65}{0}{2.8}
{cYellow}
{IDAFT+FD\\Zero Padding}
{$\mathbf{I}_{\mathbf K}
 \otimes \mathbf{Q}_{P}$}

\flagblock[-0.8mm]{proto}
{10.75}{0}{2.8}
{cGreen}
{Prototype\\Filter Matrix}
{$\mathbf{G}$}

\flagblock{hpa}
{14.05}{0}{2.9}
{cOrange}
{High Power\\Amplifier}
{$\mathbf{y}
=\mathbf{g}(\mathbf{s})
=\kappa\mathbf{s}+\mathbf{d}$}


\flagblock{symbol}
{7.25}{-3.50}{2.8}
{cLime}
{Symbol\\Estimation}
{$\hat{x}$}

\flagblock{detect}
{10.85}{-3.50}{3.2}
{cLime}
{Detector}
{\scriptsize LMMSE, GaBP, FOGa}

\node[
    font=\small
] at (-0.35,0.42)
{$x\in\mathcal{D}^{Z}$};


\draw[->]
(-0.35,0) -- (0.35,0);

\draw[->]
(2.75,0) -- (3.15,0);

\draw[->]
(5.95,0) -- (6.25,0);

\draw[->]
(9.05,0) -- (9.35,0);

\draw[->]
(12.15,0) -- (12.60,0)
node[
    midway,
    above,
    font=\small
]
{$\mathbf{s}$};

\draw[->]
(15.50,0) -- (16.15,0)
node[
    midway,
    above,
    font=\small
]
{$\mathbf{y}$};

\node[
    draw=black,
    fill=cCyan,
    rounded corners=4pt,
    line width=0.5pt,
    align=center,
    font=\scriptsize,
    minimum width=2.45cm,
    minimum height=0.75cm,
    inner sep=2pt
] (ambig) at (14.15,-1.65)
{\textbf{Ambiguity Function}\\
$\mathcal{A}_{\mathbf y}(l,f)$};

\draw[->]
(15.75,0)
|- (ambig.east);

\node[
    font=\scriptsize\itshape,
    anchor=south
] at (14.15,-1.30)
{Waveform metric};


\draw[->]
(16.15,-3.50) -- (12.45,-3.50)
node[
    midway,
    above,
    font=\small
]
{$\mathbf{r}
=\mathbf{H}\mathbf{y}
+\mathbf{n}$}
node[
    midway,
    below,
    font=\small
]
{Received Signal};

\draw[->]
(9.25,-3.50) -- (8.65,-3.50);

\boxtitle{0.27}{0.98}
{Transmitter}

\boxtitle{5.72}{-2.52}
{Communication Receiver}

\end{scope}
\end{tikzpicture}%
        }%
        \hspace*{13mm}%
    }
    \vspace{-2ex}
    \caption{Schematic of the \ac{AFBM} transmitter, sensing and communications receiver.}
    \label{fig:AFBMmod_schematic}
\end{figure*}
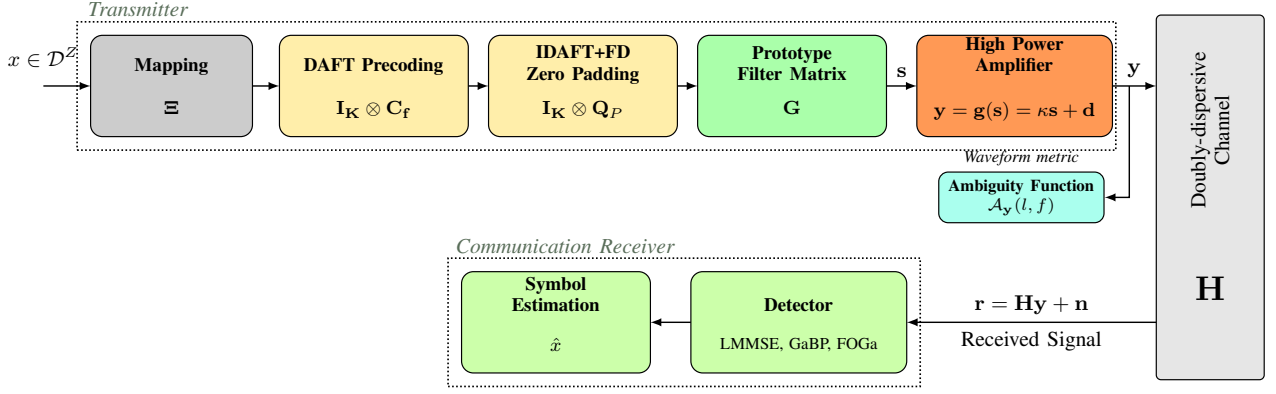

\subsection{Transmit Signal Model}

The considered system model is presented in Figure \ref{fig:AFBMmod_schematic}. 
As a starting point, let $L$ denote the number of active subcarriers in an \ac{AFBM} system with $N$ total subcarriers. The system is organized into blocks of $K$ symbols, each with a duration of $T/2$ seconds, to keep the same rate as an AFDM system transmitting for each $T$ seconds. Each subcarrier is spaced by $F$ Hz. This results in a time-frequency grid with $L/2$ points in frequency, spaced by $F$ Hz, and $K$ points in time, spaced by $T/2$ seconds. Accordingly, the total bandwidth is given by $B = L F$, and the total transmission interval by $K T/2$. 

$\mathbf{x} \in \mathcal{D}^Z$ denotes the complex transmit symbol vector for the proposed scheme, where $\mathcal{D}$ denotes the complex modulation constellation and $Z \triangleq {KL/2}$. For the sake of convenience, $L = N/2$ is adopted.
Symbols in $\mathbf{x}$ are arranged in the first and last $L/4$ positions of a matrix $\mathbf{A} \in \mathbb{C}^{L \times K}$ to uphold a quasicomplex orthogonality condition~\cite{ranasinghe2025affinefilterbankmodulation}. This mapping $\bm{\Xi}$ is expressed as
\begin{equation}
\label{eq:positions}
\mathbf{a} \triangleq \mathrm{vec}(\mathbf{A}) = \bm{\Xi} \mathbf{x} \in \mathbb{C}^{LK \times 1},
\end{equation}
where $\mathrm{vec}(\cdot)$ denotes the column-wise vectorization operation and $\bm{\Xi} \in \mathbb{C}^{LK \times Z}$ is defined as $
\label{eq:Xi}
\bm{\Xi} \triangleq \mathbf{I}_K \otimes \bar{\bm{\Xi}}$, with $\bar{\bm{\Xi}} \in \mathbb{C}^{L \times \frac{L}{2}}$ given by
\begin{equation}
\bar{\bm{\Xi}} \triangleq 
\begin{bmatrix}
\mathbf{I}_{L/4} & \mathbf{0}_{L/4} \\
\mathbf{0}_{L/2 \times L/4} & \mathbf{0}_{L/2 \times L/4} \\
\mathbf{0}_{L/4} & \mathbf{I}_{L/4}  
\end{bmatrix}.
\end{equation}

Overall, the complete \ac{AFBM} transmit signal in the \ac{TD} for all $K$ blocks $\mathbf{s} \in \mathbb{C}^{M \times 1}$, with $M \triangleq ON + \tfrac{N}{2}(K-1)$, can be written as
\begin{align}
\label{eq:td_tx_signal}
\mathbf{s} &= \mathbf{G} \big(\mathbf{I}_{K} \otimes \mathbf{Q}_{P}\big) \cdot \big(\mathbf{I}_K \otimes \mathbf{C}_f \big) \mathbf{a} \nonumber \\
& = \mathbf{G} \big(\mathbf{I}_{K} \otimes \mathbf{Q}_{P} \mathbf{C}_f \big) \bm{\Xi} \mathbf{x} \nonumber \\ &= \bar{\mathbf{G}} \mathbf{x},
\end{align}
where $\bar{\mathbf{G}}=\mathbf{G} \big(\mathbf{I}_{K} \otimes \mathbf{Q}_{P} \mathbf{C}_f \big) \bm{\Xi} \in \mathbb{C}^{M \times Z}$ is the overall modulation matrix. It can be expressed in terms of the precoding matrix $\mathbf{C}_f$ in \eqref{ferf44}, the modified \ac{IDAFT} matrix $\mathbf{Q}_P$ in \eqref{eq:Q_P}, and the block Toeplitz filter matrix $\mathbf{G} \in  \mathbb{R}^{M \times NK}$, defined as
\begin{equation}
\mathbf{G} = 
\scalemath{0.8}{\begin{bmatrix}
\mathbf{G}_0  & \mathbf{0} & \mathbf{0} & \mathbf{0} & \ldots  & \mathbf{0} \\
\mathbf{0} & \mathbf{G}_1  & \mathbf{G}_0  & \mathbf{0} & \ldots  & \mathbf{0} \\
\mathbf{G}_2  & \mathbf{0} & \mathbf{0} & \mathbf{G}_1  &  \ldots  & \mathbf{0} \\
\mathbf{0} & \mathbf{G}_3  & \mathbf{G}_2 & \mathbf{0} & \ldots  & \mathbf{0} \\
\vdots &  \mathbf{0} &  \mathbf{0} & \mathbf{G}_3  &  \ldots  & \mathbf{0} \\
\vdots &  \vdots &  \vdots &  \vdots &  \ddots  &   \vdots \\
\mathbf{G}_{2O-4} & \vdots &  \vdots &  \vdots & \ddots  &  \mathbf{0} \\
\mathbf{0} & \mathbf{G}_{2O-3} & \mathbf{G}_{2O-4} &  \vdots & \ddots & \mathbf{G}_1 \\
\mathbf{G}_{2O-2} &  \mathbf{0} &  \mathbf{0} &  \mathbf{G}_{2O-3} & \ddots &  \mathbf{0} \\
\mathbf{0} & \mathbf{G}_{2O-1} & \mathbf{G}_{2O-2} & \mathbf{0} & \ddots &  \mathbf{G}_3 \\
\vdots &  \mathbf{0} & \mathbf{0} & \mathbf{G}_{2O-1} & \ddots & \vdots \\
\vdots &  \vdots & \ddots &   \vdots &  \ddots &  \mathbf{0} \\
\mathbf{0} &   \mathbf{0} &  \ldots & \mathbf{0} &  \ddots & \mathbf{G}_{2O-1} 
\end{bmatrix}},
\label{fhddg}
\end{equation}
where $\mathbf{G}_p \in  \mathbb{R}^{N/2 \times N/2}$ denote the diagonal matrix of real filter coefficients, $i.e.$,
\begin{equation}
\mathbf{G}_p = \mathrm{diag}(\mathbf{g}_p), \;\;\; p = 0,1,2,\ldots,2O-1,
\end{equation}
where
\begin{equation*}
\mathbf{g}_p = [g[pN/2], g[pN/2+1], \ldots , g[pN/2+N/2-1]],
\end{equation*}
and $\mathbf{g}$ represents the prototype filter of length $ON$, with $O$ denoting the overlap factor. 
The structure of $\mathbf{G}$, through the inclusion of $\mathbf{0}_{N/2 \times N/2}$ matrices and with the representation as a sum of delayed matrices~\cite{pereira2022generalized}, ensures that the transmitted symbols are delayed from one another every $N/2$ samples. 
%
$\mathbf{Q}_{P}$ is comprised of an \ac{IDAFT} whose output is zero-padded in the frequency domain, obtained as
\begin{equation}
\mathbf{Q}_{P} = \mathbf{F}_{N}^{H}\mathbf{T}_{NP}\mathbf{F}_P \mathbf{\tilde{W}}\herm_{P},
\label{eq:Q_P}
\end{equation}
where 
$\mathbf{T}_{NP}\triangleq \begin{bmatrix} [\mathbf{I}_{P/2}  \;  \mathbf{0}_{P/2} ]^T& \mathbf{0}_{P \times (N-P)} &[  \mathbf{0}_{P/2} \; \mathbf{I}_{P/2} ]^T \end{bmatrix}^{T}$
is an $N \times P$ matrix with  $\mathbf{T}_{NP}^{T}\mathbf{T}_{NP} = \mathbf{I}_P$. 
For a given $N$, $\mathbf{W}_{N} \in \mathbb{C}^{N \times N}$ is the $N$-point \ac{DAFT} matrix, defined as
\begin{equation}
\mathbf{W}_{N} = \mathbf{\Lambda}_{c_1,N}\mathbf{F}_{N}\mathbf{\Lambda}_{c_2,N},
\end{equation}
with
\begin{equation}
\mathbf{\Lambda}_{c_i,N} = \mathrm{diag}[e^{-j2\pi c_i (0)^2}, \dots, e^{-j2\pi c_i (N-1)^2}] \in \mathbb{C}^{N \times N}
\end{equation}
%
denoting an $N \times N$ diagonal chirp matrix with central digital frequency $c_i$. We recall that $\mathbf{F}_{N}^{H}\mathbf{T}_{NP}\mathbf{F}_P$, where $P < N$, will guarantee that the chirps are sampled at a rate lower than the Nyquist rate and introduce frequency domain zero padding, enabling frequency containment~\cite{savaux2024special}.
The extended \ac{DAFT} $\mathbf{\tilde{W}}_{P} \in  \mathbb{C}^{L \times P}$ 
is defined as
\begin{equation}
\mathbf{\tilde{W}}_{P} = 
\begin{bmatrix}
\mathbf{I}_L &  \mathbf{0}_{L\times (P-L)}  
\end{bmatrix}
\mathbf{W}_P.
\label{dft_espalhada}
\end{equation}

Finally, let us define  $\mathbf{C}_f \in \mathbb{C}^{L \times L}$ as
%
%
\begin{equation}
\mathbf{C}_f \triangleq \mathbf{W}_{L} \mathrm{diag}({{\mathbf{\tilde{b}}}}),
\label{ferf44}
\end{equation}
where the $\tilde{l}$-th element of $\tilde{\mathbf{b}}$ is obtained as
\begin{equation*}
\tilde{b}({\tilde{l}}) = 
\begin{cases} 
\sqrt{\frac{1}{{\tilde{c}}({\tilde{l}})}}, & \tilde{l} \in \left[ 0,\ldots,\tfrac{L}{4}-1 \right] \cup \left[ L-\tfrac{L}{4},\ldots,L-1 \right] \\[1ex]
0, & \text{otherwise},
\end{cases}
\end{equation*}
with
\begin{equation*}
\mathbf{\tilde{c}} \triangleq \mathrm{diag}(\mathbf{W}\herm_L\mathbf{Q}_{P}^{H}\mathbf{\widetilde{G}}^H\mathbf{\widetilde{G}}\mathbf{Q}_{P}\mathbf{W}_L), 
\end{equation*}
and $\widetilde{\mathbf G} \in \mathbb{R}^{ON\times N}$ is the single-symbol filtering matrix constructed from $\mathbf{G}_p$~\cite{ranasinghe2025affinefilterbankmodulation}. Based on the proposed system model, the waveform can be interpreted as a filtered version of the DAFT-spread \ac{AFDM} scheme~\cite{tao2025affine}, where the standard sinc-chirp subcarriers are replaced with chirp-filtered subcarriers. Here, the considered filter is well localized both in time and in frequency and implemented with a filterbank. Due to this well-localized filter, a \ac{CP} is not used in \ac{AFBM}.

\subsection{Modeling of the \ac{HPA} Nonlinearities}
In this paper, we consider a memoryless \ac{HPA} model, as commonly assumed when the \ac{HPA} bandwidth is significantly larger than that of the transmitted signal, which is the case in most commercial communication systems~\cite{guel2009etude}. Moreover, memory effects correspond to frequency-selective behavior and are assumed to be compensated for at the transmitter.

\subsubsection{Rapp Model}
The Rapp model is a memoryless, widely accepted \ac{SSPA} model encompassing amplitude clipping ($i.e.$, AM-AM distortion)~\cite{rapp1991effects,guel2009etude}. The $n$-th element of the amplified signal can be written as
\begin{equation} y(n) = g(s(n)) =s(n)\left [{1 + \left ({\frac {|s(n)|}{A_{sat}}}\right) ^{2q }}\right ]^{-\frac {1}{2q }},  \end{equation} 
where $g(\cdot)$ is the \ac{HPA} transfer function and $q$ is the smoothness factor that controls the transition from the linear to the saturation domain determined by the input saturation voltage $A_{sat}$.
In practice, to mitigate the impacts of the nonlinear distortion, the \ac{HPA} operates at an \ac{IBO} from a given \ac{HPA} operating point. The \ac{IBO}, usually expressed in dB, is defined as the ratio between the 1-dB compression input power $\mathcal{P}_{in,\mathrm{1dB}}$ and the average input signal power 
\begin{equation}
\mathrm{IBO}
=
10\log_{10}\left(
\frac{\mathcal{P}_{in,\mathrm{1dB}}}{\alpha ^2\sigma_s^2}
\right)
\quad (\mathrm{dB}),
\end{equation}
where  $\sigma_s^2=\mathbb{E}[|s(n)|^2]$ denotes the mean input signal power, and the back-off coefficient $\alpha \in (0,1]$ is
\begin{equation}
\alpha = \sqrt{\frac{\mathcal{P}_{in,\text{1dB}}}{ \sigma_s^2 } 10^{- \text{IBO}/10}}.
\end{equation}

\subsubsection{Bussgang Decomposition}
Since the input of the \ac{HPA} is approximately Gaussian~\cite{gourar2026robustness}, the output of a memoryless nonlinear \ac{HPA} can be expressed using the Bussgang decomposition~\cite{bussgang1952crosscorrelation,demir2021bussgang} as
\begin{equation}
\mathbf{y} = \kappa \mathbf{s} + \mathbf{d} =  \kappa \bar{\mathbf{G}}  \mathbf{x} + \mathbf{d},
\label{eq:y_sspa}
\end{equation}
where $\mathbf{d}$ denotes the nonlinear distortion with variance $\sigma_d^2$, uncorrelated with the input signal $\mathbf{s}$, and $\kappa$ is the complex Bussgang gain given by
\begin{equation}
\kappa = \frac{\mathbb{E}[\mathbf{s}^{H}g(\mathbf{s})]}{\mathbb{E}[\mathbf{s}^H\mathbf{s}]}.
\label{eq:kappa_afbm}
\end{equation}

The variance $\sigma_d^2$ of the nonlinear distortion $\mathbf{d}$ is given by
\begin{equation}
    \sigma_d^2  = \mathbb{E}\big[|d(n)|^2\big] = \mathbb{E}\big[|g(s(n))|^2\big] - |\kappa|^2 \mathbb{E}\big[|s(n)|^2\big].
    \label{eq:sigmad_afbm}
\end{equation}

\section{Analysis of the Ambiguity Function} \label{sec:af_analysis}
To explore the inherent characteristics of the communication waveforms in terms of sensing capabilities, we study the \ac{AF}.
\subsection{Definition}
For a discrete-time signal $\mathbf{s}$, the \ac{AF} is defined as
\begin{equation}
\mathcal{A}(l,f) = \sum_{n=1}^M s(n+l) s^\ast(n) e^{-j2\pi f n/M},
\label{eq:AF}
\end{equation}
where $l$ denotes the delay index and $f$ the Doppler frequency index. In practice, the delay index spans $l\in[-(M-1),M-1]$, corresponding to all possible linear shifts of an $M$-sample sequence, while the Doppler frequency lies in $f\in[-M/2,M/2]$~\cite{rou2025normalized}.
Using vector notation, the \ac{AF} can be expressed in matrix form as~\cite{bedeer2025ambiguity}
\begin{equation}
\mathcal{A}(l,f) =
\mathbf{s}^H \mathbf{D}^{f} \mathbf{J}_l \mathbf{s},
\label{eq:AF_matrix}
\end{equation}
where $\mathbf{J}_l$ is the $l$-samples linear left shift matrix whose elements are defined as
\begin{equation} \mathbf{J}_l = 
\begin{bmatrix} \mathbf{0}_{(M-l)\times l} & \mathbf{I}_{M-l} 
\\ \mathbf{0}_{l\times l} & \mathbf{0}_{l\times (M-l)} 
\end{bmatrix},
\label{eq:delay-shift-matrix}
\end{equation}
and $\mathbf{D} \in \mathbb{C}^{M \times M}$ is the diagonal roots-of-unity matrix, given by
\begin{equation}
\mathbf{D}
=
\operatorname{diag}
\!\left(
\{e^{-j2\pi k/M}\}_{k=0}^{M-1}
\right), \label{eq:Doppler-shift-matrix}
\end{equation}
and is raised to the power $f$. 
Substituting the signal model in \eqref{eq:td_tx_signal} into \eqref{eq:AF_matrix}, the \ac{AF} can be written compactly as
\begin{equation}
\mathcal{A}(l,f)
=
\mathbf{x}^H
\mathbf{\Phi}_{l,f}
\mathbf{x},
\label{eq:AF_qaudratic}
\end{equation}
where $
\mathbf{\Phi}_{l,f} =\bar{\mathbf{G}}^H \mathbf{D}^f \mathbf{J}_l \bar{\mathbf{G}}$ is the modulation ambiguity matrix.

\subsection{Preliminaries: Properties of the Ambiguity Function}

\textbf{{Property 1}} (Sensitivity to the constellation symbols)\\
To evaluate the impact of constellation symbols statistics on the sidelobes, we study the expected squared magnitude of the AF, expressed as~\cite{bedeer2025ambiguity}
\begin{align}
\mathbb{E}\big[|\mathcal{A}(l,f)|^2\big]
&= \sigma_x^4\Big(|\operatorname{tr}(\mathbf{\Phi}_{l,f})|^2 + \|\mathbf{\Phi}_{l,f}\|_F^2\Big) \nonumber \\
&\quad + (\mu_4 - 2\sigma_x^4 )\| \operatorname{diag}(\mathbf{\Phi}_{l,f}) \|^2_2,
\label{AF_squared}
\end{align}
where $\sigma_x^2=\mathbb{E}[|x(k)|^2]$, $|\operatorname{tr}(\mathbf{\Phi}_{l,f})|^2  = \!\sum_{i,k} \mathbf{\Phi}_{l,f}(i,i)\,\mathbf{\Phi}_{l,f}^\ast(k,k) $, $\|\mathbf{\Phi}_{l,f}\|_F^2 = \sum_{i,j} | \mathbf{\Phi}_{l,f}(i,j)|^2 $ is the Frobenius norm of the matrix $\mathbf{\Phi}_{l,f}$, and $\| \operatorname{diag}(\mathbf{\Phi}_{l,f}) \|^2_2=\sum_{n} |\mathbf{\Phi}_{l,f}(n,n)|^2$.\\

The constellation symbols' implications are generally studied for the ranging or delay sidelobes \textit{i.e,} $f=0$~\cite{liu2025cp}, given that the randomness of the data payload directly impacts the autocorrelation properties.
The dependence of the sidelobe energy on constellation statistics enters through the term $(\mu_4-2\sigma_x^4)$. For PSK constellations, $\mu_4=1$, making this term constant. For \ac{QAM} constellations, $\mu_4$ varies only slightly (see Table~\ref{tab:kurtosis}), leading to marginal variations in ambiguity sidelobe levels, but also mainly if $\|\mathbf{\Phi}_{l,0}\|_F^2 \gg \| \operatorname{diag}(\mathbf{\Phi}_{l,0}) \|^2_2$.
\begin{table}[t]
\caption{Kurtosis values of typical constellations \cite{liu2025cp}}
\label{tab:kurtosis}
\centering
\resizebox{\columnwidth}{!}{
\begin{tabular}{c|c|c|c|c|c}
\hline
\textbf{Constellation} & PSK & 16-QAM & 64-QAM & 128-QAM & 256-QAM \\
\hline
\textbf{Kurtosis} & 1 & 1.32 & 1.381 & 1.3427 & 1.3953 \\
\hline
\end{tabular}
}
\end{table}
This depends on how the energy of $\mathbf{\Phi}_{l,0}$ is distributed between its diagonal and off-diagonal elements. If $\mathbf{\Phi}_{l,0}$ is diagonal with non-negligible diagonal terms, the constellation-dependent term has a substantial contribution. In contrast, if most of the energy lies in off-diagonal elements, the ambiguity sidelobe energy becomes invariant to the constellation choice. For the AFBM waveform, the matrix $\mathbf{\Phi}_{l,0}=\bar{\mathbf{G}}^H\mathbf J_l\bar{\mathbf{G}}$ is sparse and non-diagonal, given $\mathbf J_l$ is not diagonalized by the precoding matrix $\bar{\mathbf{G}}$.
This also implies that, for the AFBM, the difference in the ranging sidelobe energy between constant-modulus and non-constant-modulus constellations is not as substantial either, since the difference in kurtosis is on the order of $10^{-1}$, in contrast to the case of the OFDM signal, where constant-modulus constellations result in optimally low ranging sidelobes~\cite{liu2025cp}.

\textbf{{Property 2}} (Rician distribution of $|\mathcal{A}(l,f)|$)

\noindent To evaluate the sensing capability of a waveform, we derive an approximate distribution for $|\mathcal{A}(l,f)|$.
For sufficiently large $Z$, the \ac{AF} in~\eqref{eq:AF_qaudratic} is approximated as a complex Gaussian random variable using the \ac{CLT},
\begin{equation}
\mathcal{A}(l,f)
\sim
\mathcal{CN}
\!\left(
\mu(l,f),
\Sigma^2(l,f)
\right),
\end{equation}
with $\mu(l,f) = \mathbb{E}[\mathcal{A}(l,f)]$, and $\Sigma^2(l,f)=\operatorname{Var}\!\left(\mathcal{A}(l,f)\right)$. In the following, we derive $\mu(l,f)$ and $\Sigma(l,f)$. 

Using $\mathbb{E}\left[\mathbf{x}^H \mathbf{\Phi}_{l,f} \mathbf{x}\right] \!
= \! \operatorname{tr}\left(\mathbf{\Phi}_{l,f}\mathbb{E}[\mathbf{x}\mathbf{x}^H] \right)$ and $\mathbb{E}[\mathbf{x}\mathbf{x}^H] = \! \sigma_x^2 \mathbf I_{Z}$, we express the expectation of the \ac{AF} as follows,
\begin{align}
\mu(l,f) = \mathbb{E}[\mathcal{A}(l,f)]
&=
\sigma_x^2\operatorname{tr}
\!\left(
\mathbf{\Phi}_{l,f}
\right) .
\label{eq:mu}
\end{align}

\noindent Combining the expression of the \ac{AF} average squared magnitude~\eqref{eq:AF_qaudratic} and the mean~\eqref{eq:mu}, the \ac{AF} variance satisfies
\begin{align}
\Sigma^2(l, f)&= \mathbb{E}[|\mathcal{A}(l,f)|^2] - |\mu(l,f)|^2  \nonumber \\
\quad &=  \sigma_x^4\|\mathbf{\Phi}_{l,f}\|_F^2 + (\mu_4 - 2\sigma_x^4 )\sum_{k=0}^{Z-1} |\mathbf{\Phi}_{l,f}(k,k)|^2.
\label{Var}
\end{align}

Since $\mathcal{A}(l,f)$ is approximately complex Gaussian, its magnitude follows a Rician distribution~\cite{bedeer2025ambiguity}
\begin{equation} |\mathcal{A}(l,f)| \sim \text{Rice} \left( |\mu(l,f)|, \frac{\Sigma^2(l,f)}{2} \right). \end{equation}

\begin{figure}[t]
  \centering
    \subfloat[Zero-Doppler cut]{%
  \includegraphics[width=1\linewidth]{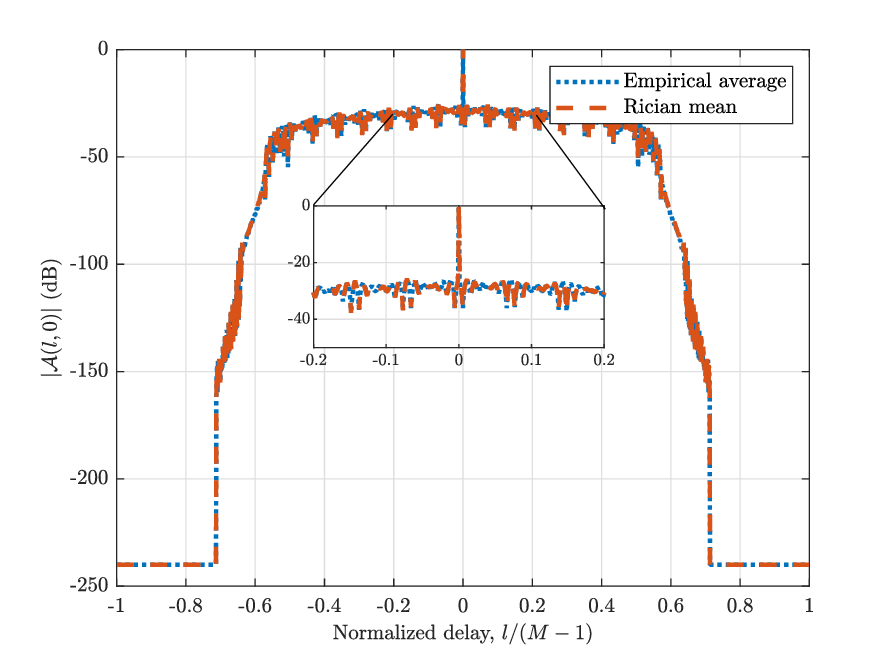}%
    }\\
    \subfloat[Zero-delay cut]{%
  \includegraphics[width=1\linewidth]{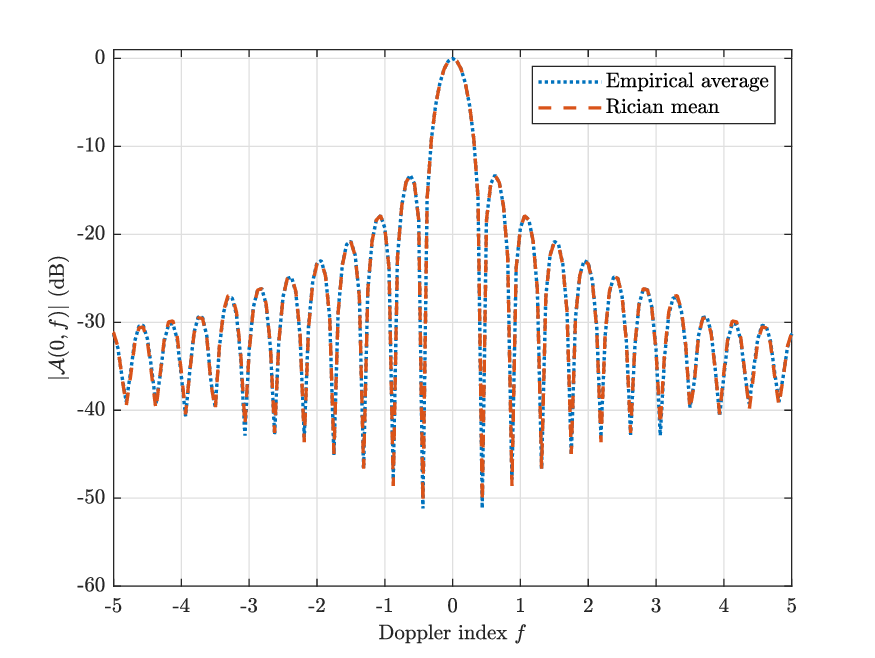}%
    }
    \caption{Comparison of empirical average and Rice distribution of the magnitude of (a) Zero-Doppler and (b) Zero-delay cuts, $L = 64$, $N=128$, $K=8$, and $O=1.5$, with the Hermite prototype filter~\cite{haas1997time}.}
    \label{fig:linear_rician}
\end{figure}

Fig~\ref{fig:linear_rician} shows the empirical average and the approximate Rice distribution of the magnitude of the AF, and they closely match, validating our derivations.

\subsection{Ambiguity Function under \ac{HPA} Nonlinearities} \label{subsection:AF_under_PA_NL}
Let $\mathbf P_{\bar G}
\triangleq
\bar{\mathbf G}
\left(
\bar{\mathbf G}^H\bar{\mathbf G}
\right)^{-1}
\bar{\mathbf G}^H $ denote the orthogonal projector onto the modulation subspace spanned by
the columns of $\bar{\mathbf G}$. The TD distortion can be decomposed as
\begin{equation}
\mathbf d
=
\mathbf d_{\parallel}
+
\mathbf d_{\perp},
\end{equation}
where $\mathbf d_{\parallel} = \mathbf P_{\bar G}\mathbf d = \bar{\mathbf G}\mathbf t,$ and $
\mathbf d_{\perp} =\left( \mathbf I_M-\mathbf P_{\bar G} \right)\mathbf d$. The least-squares modulation-domain representation of the distortion is
therefore
\begin{equation}
\mathbf t
=
\bar{\mathbf G}^{\dagger}\mathbf d
=
\left(
\bar{\mathbf G}^H\bar{\mathbf G}
\right)^{-1}
\bar{\mathbf G}^H\mathbf d.
\label{eq:t_ls_projection}
\end{equation}
The Bussgang decomposition becomes
\begin{equation}
\mathbf y
=
\kappa \bar{\mathbf G}\mathbf x
+
\bar{\mathbf G}\mathbf t
+
\mathbf d_{\perp}.
\end{equation}
Numerical evaluation indicates that $\eta = 97\%$ of the distortion energy lies in the modulation subspace ($i.e.,$ $\mathbb{E}\left[\Vert{}\mathbf{d}_{\parallel}\Vert{}_2^2\right] \gg \mathbb{E}\left[\Vert{}\mathbf{d}_{\perp}\Vert{}_2^2\right]$), allowing us to neglect $\mathbf{d}_{\perp}$ and approximate $\mathbf{d} \approx \bar{\mathbf{G}}\mathbf{t}$. Accordingly, the amplified signal is approximated as
\begin{equation}
\mathbf y \approx \bar{\mathbf G}(\kappa\mathbf x+\mathbf t).
\label{eq:y_mod_domain}
\end{equation}

Using \eqref{eq:y_mod_domain}, the \ac{AF} of the amplified signal becomes
\begin{align}
\mathcal{A}_y(l,f)
&\approx |\kappa|^2\,\mathbf{x}^H\mathbf{\Phi}_{l,f}\mathbf{x}
+\kappa\,\mathbf{x}^H\mathbf{\Phi}_{l,f}\mathbf{t}
+\kappa^{\ast}\,\mathbf{t}^H\mathbf{\Phi}_{l,f}\mathbf{x} \nonumber \\ 
& \quad+\mathbf{t}^H\mathbf{\Phi}_{l,f}\mathbf{t}. 
\label{eq:chi_y}
\end{align}

The post-amplification \ac{AF} depends on the complete AFBM modulation chain and, in particular, on the selected prototype filter, making an exact characterization of its mainlobe and sidelobe behavior likewise analytically filter-dependent. We therefore adopt an approximate statistical characterization through its mean $\mu_y(l,f)$ and variance $\Sigma_y^2(l,f)$, which respectively capture the average levels and fluctuations of the \ac{AF} under \ac{HPA} nonlinearities.

We start by deriving the mean of $|\mathcal{A}(l,f)|$ by taking the expectation of the \ac{AF} expressed with the Bussgang parameters (in~\eqref{eq:chi_y}). Under the adopted modulation-domain approximation, we further assume that the projected distortion is uncorrelated with the transmitted symbol vector, $i.e.$, $\mathbb{E}\!\left[\mathbf{x}\mathbf{t}^{H}\right] = \mathbf{0}_Z$. It then follows that the cross $\mathbb{E}\!\left[\mathbf{x}^{H}\mathbf{\Phi}_{l,f}\mathbf{t}\right]$ and $\mathbb{E}\!\left[\mathbf{t}^{H}\mathbf{\Phi}_{l,f}\mathbf{x}\right]$ vanish. We note that the expectation of the remaining distortion term is $\mathbb{E}\left[\boldsymbol{\mathbf{t}}^H \, \mathbf{\Phi}_{l,f} \boldsymbol{\mathbf{t}}\right] \!
= \! \operatorname{tr}\left(\mathbf{\Phi}_{l,f}\mathbb{E}[\boldsymbol{\mathbf{t}}\boldsymbol{\mathbf{t}}^H ] \right) = \operatorname{tr}\left(\mathbf{\Phi}_{l,f}\mathbf{R}_{{\mathbf{t}}}\right)$, where $\mathbf{R}_{{\mathbf{t}}} = \mathbb{E}[\boldsymbol{\mathbf{t}}\boldsymbol{\mathbf{t}}^H ] $. Hence, we write the expectation of $\mathcal{A}_y(l,f)$ as follows,
\begin{align}
\mu_y(l,f)
= \mathbb{E}[\mathcal{A}_y(l,f)] = |\kappa|^2  \sigma_x^2
\operatorname{tr} \!\left(\mathbf{\Phi}_{l,f}\right) 
+  \operatorname{tr}\!\left(\mathbf{\Phi}_{l,f} \, \mathbf{R}_{{\mathbf{t}}} \,\right).
\label{eq:mu_y}
\end{align}

Next, we note that the variance is given by
\begin{align}
\Sigma^2_y(l, f)&= \mathbb{E}[|\mathcal{A}_y(l,f)|^2] - |\mu_y(l,f)|^2.
\label{eq:Var_y}
\end{align} 

We recall that the approximation of $\mathbb{E}\!\left[|\mathcal{A}_y(l,f)|^2\right]$ derived in~\cite{gourar2026robustness} is given by
\begin{align}
\mathbb E\!\left[|\mathcal A_y(l,f)|^2\right]
&\approx |\kappa|^4
\Big[
\sigma_x^4\Big(
|\operatorname{tr}(\mathbf\Phi_{l,f})|^2
+
\|\mathbf\Phi_{l,f}\|_F^2
\Big) \nonumber \\
&\quad+ 
(\mu_4-2\sigma_x^4)\|
\operatorname{diag}(\mathbf\Phi_{l,f})
\|_2^2
\Big] \nonumber \\
&\quad +
|\kappa|^2
\operatorname{tr}\!\left(
\mathbf\Phi_{l,f}\, \mathbf{R}_{{\mathbf{t}}} \,\mathbf\Phi_{l,f}^H
\right) \nonumber \\
&\quad+ 
|\kappa|^2
\operatorname{tr}\!\left(
\mathbf\Phi_{l,f}^H\, \mathbf{R}_{{\mathbf{t}}} \,\mathbf\Phi_{l,f}
\right) \nonumber \\
&\quad  + 
\left|
\operatorname{tr}\!\left(
\mathbf\Phi_{l,f}\, \mathbf{R}_{{\mathbf{t}}} \,
\right)
\right|^2 \nonumber \\
&\quad
+
\operatorname{tr}\!\left(
\mathbf\Phi_{l,f}\, \mathbf{R}_{{\mathbf{t}}} \,\mathbf\Phi_{l,f}^H\, \mathbf{R}_{{\mathbf{t}}} \,
\right)  \nonumber \\
&\quad +
2|\kappa|^2
\Re\!\left\{
\operatorname{tr}(\mathbf\Phi_{l,f})
\operatorname{tr}\!\left(
\mathbf\Phi_{l,f}\, \mathbf{R}_{{\mathbf{t}}} \,
\right)^\ast
\right\}.
\label{eq:E[|A(l,nu)|^2]}
\end{align}

Given that $
|\mu_y(l,f)|^2
=
|\kappa|^4
\left|
\operatorname{tr}
(\mathbf{\Phi}_{l,f})
\right|^2
+
\left|
\operatorname{tr}
\!\left(
\mathbf{\Phi}_{l,f}\, \mathbf{R}_{{\mathbf{t}}} \,
\right)
\right|^2 
+
2|\kappa|^2
\Re\!\left\{
\operatorname{tr}(\mathbf{\Phi}_{l,f})
\operatorname{tr}
\!\left(
\mathbf{\Phi}_{l,f}\, \mathbf{R}_{{\mathbf{t}}} \,
\right)^\ast
\right\}$, substituting \eqref{eq:E[|A(l,nu)|^2]} in~\eqref{eq:Var_y} yields
\begin{align}
\Sigma_y^2(l,f)
&\approx
|\kappa|^4
\left[
\sigma_x^4\|\mathbf{\Phi}_{l,f}\|_F^2
+
(\mu_4-2\sigma_x^4)
\left\|
\operatorname{diag}(\mathbf{\Phi}_{l,f})
\right\|_2^2
\right]
\nonumber\\
&\quad
+
|\kappa|^2
\operatorname{tr}
\!\left(
\mathbf{\Phi}_{l,f}
\, \mathbf{R}_{{\mathbf{t}}} \,
\mathbf{\Phi}_{l,f}^H
\right)
+
|\kappa|^2
\operatorname{tr}
\!\left(
\mathbf{\Phi}_{l,f}^H
\, \mathbf{R}_{{\mathbf{t}}} \,
\mathbf{\Phi}_{l,f}
\right) \nonumber \\
&\quad + 
\operatorname{tr}
\!\left(
\mathbf{\Phi}_{l,f}
\, \mathbf{R}_{{\mathbf{t}}} \,
\mathbf{\Phi}_{l,f}^H
\, \mathbf{R}_{{\mathbf{t}}} \,
\right).
\label{eq:variance_y}
\end{align}


\noindent Following the same reasoning as in the linear case, $\mathcal{A}_y(l,f)$ is approximately complex Gaussian distributed,
\begin{equation}
\mathcal{A}_y(l,f)
\sim
\mathcal{CN}
\!\left(
\mu_y(l,f),
\Sigma_y^2(l,f)
\right).
\end{equation}

Therefore, its magnitude follows a Rician distribution,
\begin{equation}
|\mathcal{A}_y(l,f)|
\sim
\text{Rice}
\left(
|\mu_y(l,f)|,
\frac{\Sigma_y^2(l,f)}{2}
\right).
\end{equation}

\begin{figure}[t]
  \centering
    \subfloat[Zero-Doppler cut]{%
  \includegraphics[width=1\linewidth]{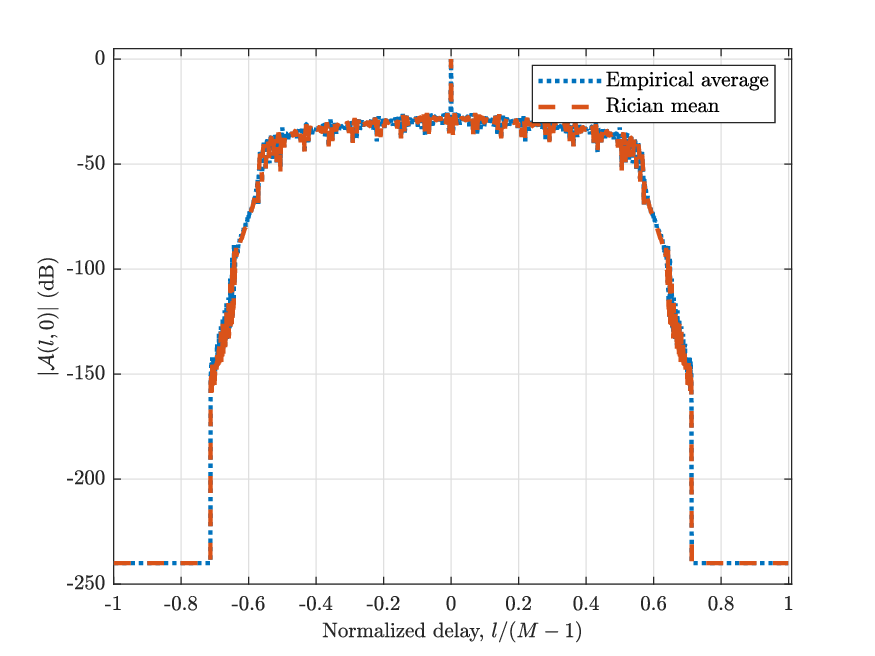}%
    }\\
    \subfloat[Zero-delay cut]{%
  \includegraphics[width=1\linewidth]{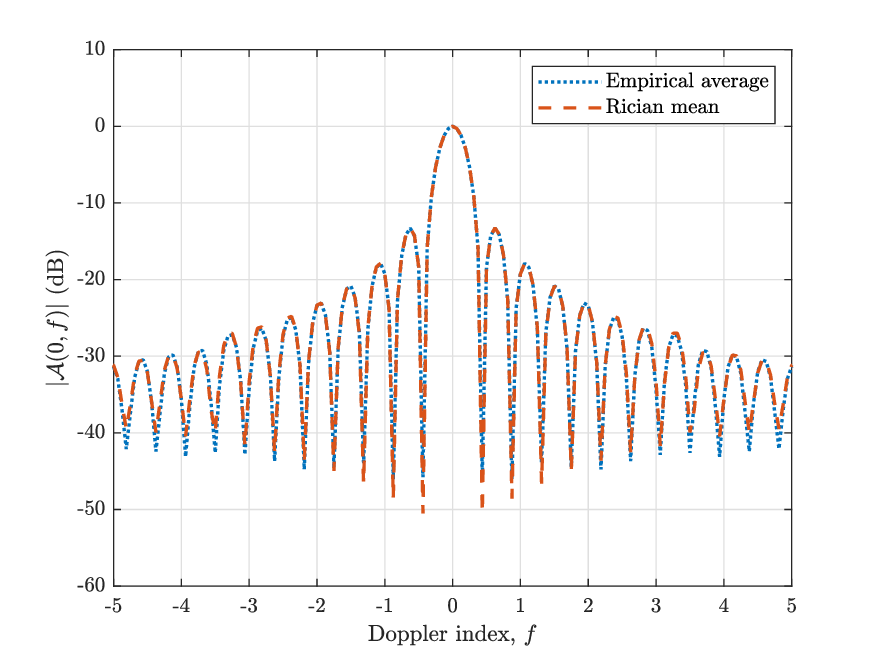}}
    \caption{Comparison of empirical average and Rice distribution of the magnitude of the (a) zero-Doppler cut and (b) zero-delay cut under \ac{HPA} nonlinearities, $A_{\text{sat}} = 1$ and $q=1.1$. $L = 64$, $N=128$, $K=8$, and $O=1.5$, with the Hermite prototype filter. }
    \label{fig:nonlinear_rician}
\end{figure}
We present in Fig.~\ref{fig:nonlinear_rician} the approximate Rice distribution of the magnitude of the zero-Doppler and zero-delay cuts, and it closely matches the empirical average, validating our derivations.

The expressions in~\eqref{eq:variance_y} and~\eqref{eq:mu_y} are formulated in terms of the Bussgang gain $\kappa$ and the projected distortion covariance matrix $\mathbf{R}_{\bf t}$. To explicitly relate the nonlinearities' impact to the \ac{HPA} parameters rather than to the random distortion statistics, we approximate the Bussgang parameters using a polynomial \ac{HPA} model. For simplicity and analytical tractability, we adopt a third-order polynomial model~\footnote{Only odd-order terms are considered in the adopted polynomial model. The first-order term represents the linear response, and the third-order term is therefore the first nonlinear contribution.}.

Provided that $\sigma_s^2 =( \sigma_x^2 / M)
\operatorname{tr}
\left(
\bar{\mathbf{G}}\,\bar{\mathbf{G}}\,^H
\right)$, approximating the sample-dependent Bussgang gain and distortion variance by their average values yields
\begin{equation}
\kappa = 1-2\alpha_3\sigma_s^2 =
1
-
\frac{2\alpha_3}{ M}\sigma_x^2
\operatorname{tr}
\left(
\bar{\mathbf{G}}\,\bar{\mathbf{G}}\,^H
\right),
\label{eq:kappa_closed_form}
\end{equation}
and
\begin{equation}
\sigma_d^2
=
2\alpha_3^2\sigma_s^6.
\label{eq:sigma_d_closed_form}
\end{equation}

\noindent \textit{Proof:} Please refer to Appendix~\ref{proof:bussgang_poly}. \hfill $\blacksquare$ 

The expressions of $\mu_y(l,f)$ and $\Sigma_y^2(l,f)$ in~\eqref{eq:mu_y} and~\eqref{eq:variance_y} are therefore rewritten using the cubic \ac{HPA} nonlinear gain $\alpha_3$, respectively in \eqref{eq:mu_ycubic_extended} and \eqref{eq:var_y_cubic_extended}.

\begin{figure*}
\begin{align}
\mu_y(l,f)
&=
\underbrace{
\left|
1
-
\frac{2\alpha_3 \sigma_x^2}{M}
\operatorname{tr}
\left(
\bar{\mathbf{G}}\bar{\mathbf{G}}^H
\right)
\right|^2 \sigma_x^2
\operatorname{tr}\!\left(\mathbf{\Phi}_{l,f}\right)
}_{\mu_{\text{signal}}\, \triangleq \, \text{useful signal term}} +
\underbrace{
\operatorname{tr}\!\left(
\mathbf{\Phi}_{l,f}
\left(
2\alpha_3^2 \sigma_x^6
\bar{\mathbf{G}}^\dagger
\left(
\bar{\mathbf{G}}\bar{\mathbf{G}}^H
\odot
\left|
\bar{\mathbf{G}}\bar{\mathbf{G}}^H
\right|^{\odot 2}
\right)
(\bar{\mathbf{G}}^\dagger)^H
\right)
\right)
}_{\mu_{\text{dist}} \, \triangleq \, \text{distortion-induced term}} .
\label{eq:mu_ycubic_extended}
\end{align}
\end{figure*}

\begin{figure*}
\begin{align}
\Sigma_y^2(l,f)
&\approx
\left.
\begin{aligned}
&
\left|
1
-
\frac{2\alpha_3\sigma_x^2}{ M}
\operatorname{tr}
\left(
\bar{\mathbf G}\bar{\mathbf G}^H
\right)
\right|^4
\left[
\sigma_x^4\|\mathbf{\Phi}_{l,f}\|_F^2
+
(\mu_4-2\sigma_x^4)
\left\|
\operatorname{diag}
(
\mathbf{\Phi}_{l,f}
)
\right\|_2^2
\right]
\end{aligned}
\right\}
\quad 
\Sigma_{\text{signal}}^2 \triangleq \text{useful signal term}
\nonumber\\[2ex]
&\,
\left.
\begin{aligned}
&+2\alpha_3^2 \sigma_x^6
\left|
1
-
\frac{2\alpha_3\sigma_x^2}{M}
\operatorname{tr}
\left(
\bar{\mathbf G}\bar{\mathbf G}^H
\right)
\right|^2
\operatorname{tr}
\left(
\mathbf{\Phi}_{l,f}^H
\bar{\mathbf G}^{\dagger} \Big(
\bar{\mathbf{G}}\bar{\mathbf{G}}^H \odot \left| \bar{\mathbf{G}}\bar{\mathbf{G}}^H \right|^{\odot 2}\Big)
(\bar{\mathbf G}^{\dagger})^H
\mathbf{\Phi}_{l,f}
\right)
\\[1ex]
&+2\alpha_3^2 \sigma_x^6
\left|
1
-
\frac{2\alpha_3 \sigma_x^2}{M}
\operatorname{tr}
\left(
\bar{\mathbf G}\bar{\mathbf G}^H
\right)
\right|^2
\operatorname{tr}
\left(
\mathbf{\Phi}_{l,f}
\bar{\mathbf G}^\dagger \Big(
\bar{\mathbf{G}}\bar{\mathbf{G}}^H \odot \left| \bar{\mathbf{G}}\bar{\mathbf{G}}^H \right|^{\odot 2}\Big)
(\bar{\mathbf G}^{\dagger})^H
\mathbf{\Phi}_{l,f}^H
\right)
\\[1ex]
&+
4\alpha_3^4 \sigma_x^{12}
\operatorname{tr}
\left(
\mathbf{\Phi}_{l,f}
\bar{\mathbf G}^{\dagger}
\left(
\bar{\mathbf{G}}\,\bar{\mathbf{G}}\,^H
\odot
\left|
\bar{\mathbf{G}}\,\bar{\mathbf{G}}\,^H
\right|^{\odot 2}
\right)
(\bar{\mathbf G}^{\dagger})^H\,
\mathbf{\Phi}_{l,f}^H
\bar{\mathbf G}^{\dagger}
\left(
\bar{\mathbf{G}}\,\bar{\mathbf{G}}\,^H
\odot
\left|
\bar{\mathbf{G}}\,\bar{\mathbf{G}}\,^H
\right|^{\odot 2}
\right)
(\bar{\mathbf G}^{\dagger})^H
\right)
\end{aligned}
\right\}
\quad
\begin{aligned}
\Sigma_{\text{dist}}^2 \triangleq \\ \text{distortion-}\\
\text{induced}\\
\text{term}
\end{aligned}
\label{eq:var_y_cubic_extended}
\end{align}
\end{figure*}
\noindent \textit{Proof:} The covariance matrix of the projected distortion is
\begin{equation}
\mathbf R_{\mathbf t}
=
\bar{\mathbf G}^{\dagger}
\mathbf R_d
(\bar{\mathbf G}^{\dagger})^H.
\label{eq:Rt_projection}
\end{equation}
Noting that $\mathbf R_s= \mathbb{E}[\mathbf s \mathbf s^H]=\sigma_x^2\bar{\mathbf G}\bar{\mathbf G}^H$,
and that the distortion covariance satisfies
$\mathbf R_d=2\alpha_3^2
(\mathbf R_s\odot|\mathbf R_s|^{\odot2})$,
we obtain
\begin{equation}
\mathbf R_d
=
2\alpha_3^2\sigma_x^6
\left(
\bar{\mathbf G}\bar{\mathbf G}^{H}
\odot
\left|
\bar{\mathbf G}\bar{\mathbf G}^{H}
\right|^{\odot 2}
\right).
\end{equation} Hence \eqref{eq:Rt_projection} becomes
\begin{equation}
\mathbf R_{\mathbf t}
=
2\alpha_3^2\sigma_x^6
\bar{\mathbf G}^{\dagger}
\left(
\bar{\mathbf G}\bar{\mathbf G}^{H}
\odot
\left|
\bar{\mathbf G}\bar{\mathbf G}^{H}
\right|^{\odot 2}
\right)
(\bar{\mathbf G}^{\dagger})^H.
\label{eq:Rt_cubic}
\end{equation} \hfill $\blacksquare$

The analysis is restricted to operating points satisfying $2\alpha_3\sigma_s^2<1$. We can observe from \eqref{eq:mu_ycubic_extended} and \eqref{eq:var_y_cubic_extended} that for small values of $\alpha_3$, the useful signal component may dominate, and that for stronger nonlinearities $i.e.$, increasing $\alpha_3$, the distortion-induced terms become more significant and may increase the \ac{AF} statistics, depending on how the ambiguity matrix $\mathbf{\Phi}_{l,f}$ projects onto the distortion covariance. Furthermore, since we have shown that the AFBM AF characteristics are weakly dependent on the constellation statistics, the behavior in the presence of \ac{HPA} nonlinearities will likewise be independent of the particular constellation used.

In the following section, the communications aspect of the \ac{AFBM} waveform under \ac{HPA} nonlinearities will be tackled.

\section{Gaussian Approximation Belief Propagation-based Receivers} \label{sec:receivers}

The distorted signal $\mathbf{y}$ at the output of the transmitter is expressed as in \eqref{eq:y_sspa}.
Correspondingly, the signal at the input of the receiver is given by
\begin{align}
    \mathbf{r} &= \mathbf{H} \mathbf{y} + \mathbf{n} \nonumber \\
     &= \underbrace{\kappa \mathbf{H} \bar{\mathbf{G}}}_{\triangleq \mathbf{B} }  \mathbf{x} + \underbrace{\mathbf{H}\mathbf{d}  + \mathbf{n}}_{\triangleq \mathbf{w} } \nonumber \\
    &=\mathbf{B} \mathbf{x} + \mathbf{w},
    \label{syst_mod_rec_AFBM}
\end{align}
where $\mathbf{H} \in \mathbb{C}^{M \times M}$ denotes the doubly-dispersive channel composed of $R$ resolvable paths. 
Each $r$-th path induces a delay $\tau_r \in [0, \tau^\mathrm{max}]$ and a Doppler shift $\nu_r \in [-\nu^\mathrm{max}, +\nu^\mathrm{max}]$, with integer delay index $\ell_r \triangleq \lfloor \tfrac{\tau_r}{T_\mathrm{s}} \rceil \in \mathbb{N}_0$ and Doppler index $f_r \triangleq \tfrac{M\nu_r}{f_\mathrm{s}} \in \mathbb{R}$, where $f_\mathrm{s} \triangleq \tfrac{1}{T_\mathrm{s}}$ denotes the sampling frequency. 
The noise vector $\mathbf{n} \in \mathbb{C}^{M \times 1}$ represents \ac{AWGN} samples with variance $\sigma_n^2$. The overall matrix $\mathbf{B}$ is of size $M\times Z$. The variance of the noise $\mathbf{w}$ is $\sigma^2_n + \sigma^2_{d}$.
As described in~\cite{rou2024orthogonal}, the channel matrix $\mathbf{H}$ is expressed as
%
\begin{equation}
\mathbf{H} \triangleq \sum_{r=1}^{R} h_r \mathbf{D}^{f_r} \mathbf{J}_{l_r} \in \mathbb{C}^{M \times M},
\label{eq:H_channel}
\end{equation}
where $h_r \in \mathbb{C}$ denotes the complex fading coefficient of the $r$-th path, and $\mathbf{J}_{l_r}$, and $\mathbf{D}$ are defined as in \eqref{eq:delay-shift-matrix} and \eqref{eq:Doppler-shift-matrix}, respectively.
In total, $\mathbf{H}$ is formed by $R$ diagonals, with positions determined by the path delays and coefficients modulated by the Doppler shifts.

The standard receiver does not have knowledge of $\sigma^2_{d}$, while a more advanced transceiver design can take into account the knowledge of the distortion introduced by the \ac{HPA} nonlinearities. These will be introduced in the next subsection. 
%

\subsection{GaBP-based Receiver}

In~\cite{ranasinghe2025affinefilterbankmodulation}, the authors have proposed to perform the AFBM data detection using the so-called Gaussian Approximation Belief Propagation (GaBP)~\cite{5503188,8543847}. In this approach, only means and variances are exchanged between the variable nodes and the factor nodes of the factor graph associated to (\ref{syst_mod_rec_AFBM}). 

Let us denote $\mu_{x_i\rightarrow f_j}^t(x_i)$ the message sent from the variable node $x_i$ to factor node $f_j$ in the $t$-th iteration, and let us denote $\mu_{f_j\rightarrow x_i}^t(x_i)$ the message from the factor node $f_j$ to variable node $x_i$. 
The minimum Kullback-Leibler divergence criterion~\cite{minka2005divergence} is applied to calculate parameters $\hat{x}_{x_i\rightarrow f_j}^t$ (mean of projection distribution) and $\hat{\tau}_{x_i\rightarrow f_j}^t$ (variance of projection distribution). The message updated from the variable nodes to factor nodes are equal to

 \begin{equation}
    \begin{aligned}
    \hat{x}_{x_i\rightarrow f_j}^t=\sum_{x_i\in \mathcal{D}}x_i\mu_{x_i\rightarrow f_j}^t(x_i)
    \end{aligned},
    \label{gam3}
\end{equation}

 \begin{equation}
    \begin{aligned}
    \hat{\tau}_{x_i\rightarrow f_j}^t=\sum_{x_i\in \mathcal{D}}|x_i|^2\mu_{x_i\rightarrow f_j}^t(x_i)-|\hat{x}_{x_i\rightarrow f_j}^t|^2
    \end{aligned}.
     \label{gav3}
\end{equation}

Let $\mathbf{B}=\{ b_{j,i} \}_{1 \leq j \leq M, 1 \leq i \leq Z}$. The message $\mu_{f_j\rightarrow x_i}^t(x_i)$ can be  approximated by a complex Gaussian function  as follows: 

 \begin{equation}
 \small
\mu_{f_j\rightarrow x_i}^t(x_i) \approx \mathcal{N}_{\mathbb{ C}}\left(b_{j,i}x_i;z_{f_j\rightarrow x_i}^t,f_{f_j\rightarrow x_i}^t\right ),
    \label{eq:fjxiWuGA2}
\end{equation}
where  $\mathcal{N}_{\mathbb{ C}}(x; \hat{x} ; \hat{\tau}) =  (\pi \hat{\tau})^{-1} \exp ( -|x-\hat{x}|^2/ \hat{\tau})$ . Assuming the knowledge of $\sigma_n^2$  and the equivalent noise due to the HPA nonlinearities $\sigma_{d}^2$, the parameters $z_{f_j\rightarrow x_i}^t$ (mean messages from factor nodes to variable nodes) and $f_{f_j\rightarrow x_i}^t$ (variance messages from factor nodes to variable nodes) are given by
 \begin{equation}
    \begin{aligned}\\
    z_{f_j\rightarrow x_i}^t= r_j -\sum_{l \neq i}b_{j,l}\hat{x}_{x_l\rightarrow f_j}^t
    \end{aligned},
    \label{gam2}
\end{equation}
\\
 \begin{equation}
    \begin{aligned}
    f_{f_j\rightarrow x_i}^t= \sigma_n^2 + \sigma_{d}^2 + \sum_{l \neq i}|b_{j,l}|^2\hat{\tau}_{x_l\rightarrow f_j}^t.
    \end{aligned}
    \label{gav2}
\end{equation}

Consequently, messages $\mu_{x_i\rightarrow f_j}^t(x_i)$  can be normalized as follows:
\begin{equation}
    \begin{aligned}
    \mu_{x_i\rightarrow f_j}^t(x_i)= \frac{\mathcal{N}_{\mathbb{ C}}\left (x_i;\zeta_{x_i\rightarrow f_j}^{t-1},\gamma_{x_i\rightarrow f_j}^{t-1}\right )}{\sum_{x_i \in \mathcal{D}}\mathcal{N}_{\mathbb{ C}}\left (x_i;\zeta_{x_i\rightarrow f_j}^{t-1},\gamma_{x_i\rightarrow f_j}^{t-1}\right )},
    \end{aligned}
    \label{gavn2fn}
\end{equation}
where $\gamma_{x_i\rightarrow f_j}^{t-1}$ (variance messages from variable nodes to factor nodes) and $\zeta_{x_i\rightarrow f_j}^{t-1}$ (means messages from variable nodes to factor node) are given by

 \begin{equation}
    \begin{aligned}
    \gamma_{x_i\rightarrow f_j}^{t}=\left (\sum_{d \neq j}\frac{|b_{d,i}|^2}{f_{f_d\rightarrow x_i}^{t}}\right )^{-1}
    \end{aligned},
    \label{gav1}
\end{equation}

 \begin{equation}
    \begin{aligned}
    \zeta_{x_i\rightarrow f_j}^{t}= \gamma_{x_i\rightarrow f_j}^{t}\sum_{d \neq j}\frac{b_{d,i}^*z_{f_d\rightarrow x_i}^{t}}{f_{f_d\rightarrow x_i}^{t}}.
    \end{aligned}
    \label{gam1}
\end{equation}

After $T$ iterations, the marginal distribution $\mu_{x_i}^{~T}(x_i)$ can be calculated as follows:
\begin{equation}
    \begin{aligned}
    \mu^T_{x_i}(x_i) \propto  \exp \left (-{\frac{|x_i~-~\zeta_{x_i}^{T}|^2}{\gamma_{x_i}^{T}}} \right ),
    \end{aligned}
    \label{eq:llrWuGa}
\end{equation}
where $\zeta_{x_i}^{T}$ and $\gamma_{x_i}^{T}$  are the estimated mean and variance of $x_i$
 \begin{equation}
    \begin{aligned}
    \gamma_{x_i}^{T}=\left (\sum_{d=1}^M\frac{|b_{d,i}|^2}{f_{f_d\rightarrow x_i}^{T}}\right )^{-1}
    \end{aligned},
    \label{gall1}
\end{equation}

 \begin{equation}
    \begin{aligned}
    \zeta_{x_i}^{T}= \gamma_{x_i}^{t}\sum_{d=1  }^M\frac{b_{d,i}^*z_{f_d\rightarrow x_i}^{T}}{f_{f_d\rightarrow x_i}^{T}}.
    \end{aligned}
    \label{gall2}
\end{equation}

To avoid convergence to local minimum and minimize the BER, we apply damping using a factor $0<\Delta \leq 1$ as follows:
\begin{equation}
   \mu^t_{x_i \rightarrow f_j} (x_i) =(1 - \Delta)  \mu^{t-1}_{x_i \rightarrow f_j} (x_i) + \Delta
 \mu^{t}_{x_i \rightarrow f_j} (x_i).
 \label{damping}
   \end{equation}

\subsection{FOGa-based Receiver}

We propose the first-order Gaussian approximation (FOGa) receiver, which reduces the complexity of GaBP by following a strategy in which the messages are rewritten after recursive updates, and the negligible terms are omitted in the large system limit.
Following the approximation rules given in~\cite{wu2014}, we first rewrite the standard messages in  (\ref{gavn2fn}) as follows:
 \begin{equation}
    \begin{aligned}
     \mu_{x_i}^t(x_i)= \frac{\mathcal{N}_{\mathbb{ C}}\left (x_i;\zeta_{x_i}^{t-1},\gamma_{x_i}^{t-1}\right )}{\sum_{x_i \in \mathcal{D}}\mathcal{N}_{\mathbb{ C}}\left (x_i;\zeta_{x_i}^{t-1},\gamma_{x_i}^{t-1}\right )} \;,
    \end{aligned}
    \label{eq:muxiFO}
\end{equation}
where $\gamma_{x_i}^{t-1}$ (variance messages variables nodes) and $\zeta_{x_i}^{t-1}$ (means messages variables nodes)  are the messages exchanged from variables nodes to factor nodes and are approximated as follows:
\begin{equation}
    \begin{aligned}
     \gamma_{x_i}^t=\left( \sum_{d =1}^M \frac{|b_{d,i}|^2}{f_{f_d}^t} \right)^{-1},
    \end{aligned}
    \label{eq:gammaFO}
\end{equation}
\begin{equation}
    \begin{aligned}     \zeta_{x_i}^t=\hat{x}_{x_i}^t+\gamma_{x_i}^t\sum_{d =1}^M \frac{b^*_{d,i}~z_{f_d}^t}{f_{f_d}^t},
    \end{aligned}
    \label{eq:zetaFO}
\end{equation}
with $z_{f_d}^t$  and $f_{f_d}^t$ the mean and variance messages from factor nodes to variable nodes, respectively. To initiate the exchange of messages from factor nodes to variable nodes, eq. (\ref{eq:muxiFO}) is updated for all variable nodes, and then the mean and variance of the projection distribution for each symbol of the \ac{QAM} alphabet are calculated as
\begin{equation}
    \begin{aligned}
     \hat{x}_{x_i}^t =\sum_{x_i \in \mathcal{D}}x_i \mu_{x_i}^t(x_i) \;,
    \end{aligned}
    \label{eq:xhatFO}
\end{equation}
\begin{equation}
    \begin{aligned}
     \hat{\tau}_{x_i}^t =\sum_{x_i \in \mathcal{D}}|x_i|^2\mu_{x_i}^t(x_i)-|\hat{x}_{x_i}^t|^2.
    \end{aligned}
    \label{eq:tauhatFO}
\end{equation}

Then, all the means and variances of the messages exchanged from factor nodes to variable nodes are calculated as follows
\begin{equation}
     z_{f_j}^t= r_j - \sum_{l =1}^Zb_{j,l}~\hat{x}_{x_l}^t +z_{f_j}^{t-1} \frac{\sum_{l'=1}^Z\hat{\tau}_{x_{l'}}^t|b_{j,l'}|^2 }{f_{f_j}^{t-1}},
    \label{eq:zttFO}
\end{equation}
\begin{equation}
     f_{f_j}^t =\sigma_n^2+ \sigma_{d}^2+\sum_{l=1}^Z|b_{j,l}|^2~\hat{\tau}_{x_l}^t.
    \label{eq:nutFO}
\end{equation}

The pseudo-code of the proposed FOGa detection procedure is summarized in Algorithm \ref{fo_gaa}.

\begin{algorithm}[t] 
\caption{FOGa algorithm}
	\begin{algorithmic}[1]
 \STATE Initialization:
 $\zeta_{x_i}^0=0$, $f_{f_j}^0=1000$, $z_{f_j}^0=0$, $\gamma_{x_i}^0=1000$, $t=1$.
		\WHILE{$t\leq{T}$}
		\FOR {$i=1\rightarrow{Z}$}
		\STATE Calculate message $\mu^t_{x_i}(x_i)$ using Eq.\eqref{eq:muxiFO}.
		\STATE Calculate $\hat{x}_{x_i}^t$ and $\hat{\tau}_{x_i}^t$ using Eq.\eqref{eq:xhatFO} and Eq.\eqref{eq:tauhatFO}, respectively.
		\ENDFOR
		\FOR {$j=1\rightarrow{M}$}
		\STATE Calculate $f_{f_j}^t$ and $z_{f_j}^t$ using Eq.\eqref{eq:nutFO} and Eq.\eqref{eq:zttFO}, respectively.
		\ENDFOR
		\FOR {$i=1\rightarrow{Z}$}
		\STATE Calculate $\gamma_{x_i}^t$ and $\zeta_{x_i}^t$ using Eq.\eqref{eq:gammaFO} and Eq.\eqref{eq:zetaFO}, respectively.
		\STATE Damping calculation 
		\ENDFOR
		\STATE $t=t+1$.
		\ENDWHILE
		\FOR {$i=1\rightarrow{Z}$}
		\STATE Calculate the mean estimated of $x_i$ $\zeta_{x_i}^{T}$  using Eq. \eqref{eq:zetaFO}
		\ENDFOR
	\end{algorithmic}
\label{fo_gaa}
\end{algorithm}

\subsection{Complexity Analysis}

The complexity of both algorithms is evaluated in terms of floating-point operations (FLOPs) for one iteration.
It is assumed that the multiplication of two complex numbers needs six FLOPs and the multiplication of a complex number and a real number needs two FLOPs. The $\exp(\cdot)$ operation needs one FLOP.
Computing the messages $\mu_{x_i\rightarrow f_j}^t(x_i) \; \forall i, \forall j$ 
in the GaBP algorithm and  $\mu_{x_i}^t(x_i) \; \forall i$ in the FOGa algorithm require $(9|\mathcal{D}|-1)ZM$ and $(9|\mathcal{D}|-1)Z$ FLOPs, respectively.
  $\hat{x}_{x_i\rightarrow f_j}^t$ 
    and $\hat{\tau}_{x_i\rightarrow f_j}^t \; \forall i, \forall j$ in the GaBP algorithm needs $(10 |\mathcal{D}|-2)ZM$ FLOPs and $\hat{x}_{x_i}^t$ 
    and $\hat{\tau}_{x_i}^t \; \forall i$ in the FOGa algorithm needs $(10 |\mathcal{D}|-2)Z$ FLOPs.

For computing the messages
exchanged from factor nodes to variables nodes, $ z_{f_j\rightarrow x_i}^t$ and
    $f_{f_j\rightarrow x_i}^t \; \forall i,  \forall j$ in the GaBP algorithm require $13ZM$ FLOPs while $z_{f_j}^t$ and
    $f_{f_j}^t \; \forall j$ in the FOGa algorithm require $12ZM+4M$ FLOPs.  
Finally, for computing the messages
exchanged from variables nodes to factor nodes, $    \gamma_{x_i\rightarrow f_j}^{t}$ and $ \zeta_{x_i\rightarrow f_j}^{t} \; \forall i, \forall j$ in the GaBP algorithm need $18ZM - 3Z$ FLOPs, while $\gamma_{x_i}^{t}$ and $ \zeta_{x_i}^{t} \; \forall i$ in the FOGa algorithm need $10ZM + 2Z + 2M$ FLOPs.

Thus, the overall computational complexity of one iteration of the GaBP and FOGa algorithms is $19 |\mathcal{D}| ZM + 28 ZM -3Z$ and  $22 ZM + 19 Z|\mathcal{D}|- Z +6M$  FLOPs, respectively. Based on the parameters listed in Table~\ref{tab:sim_params}, the FOGa algorithm reduces the computational complexity to approximately $21\%$ and $7\%$ of that of the GaBP algorithm for 4-QAM and 16-QAM, respectively.

\section{Numerical Results} \label{sec:results}

\subsection{Common Simulation Parameters}

Unless explicitly stated otherwise, all numerical results presented in this section are obtained using the common simulation parameters summarized in Table~\ref{tab:sim_params}, and the doubly dispersive channel parameters are listed in Table~\ref{tab:channel_params}. 

\begin{table}[!t]
\centering
\caption{Common simulation parameters.}
\label{tab:sim_params}
\begin{tabular}{@{}lcl@{}}
\toprule
Parameter & Symbol & Value \\
\midrule
Number of active subcarriers & $L$ & $64$ \\
Chirp Spread Size & $P$ & $128$ \\
Total number of subcarriers & $N$ & $128$ \\
Number of multicarrier symbols & $K$ & $8$ \\
Prototype-filter overlap factor & $O$ & $1.5$ \\
First $L$-point DAFT chirp parameter
& $c_{1,L}$ & $0.1015625$ \\
Second $L$-point DAFT chirp parameter
& $c_{2,L}$ &  $7.7712\times10^{-5}$ \\
First $P$-point DAFT chirp parameter
& $c_{1,P}$ & $0.05078125$ \\
Second $P$-point DAFT chirp parameter
& $c_{2,P}$ & $1.9428\times10^{-5}$ \\
Prototype filter & -- & Hermite \\
Simulated Frames per SNR & -- & 5000 \\
\bottomrule
\end{tabular}
\end{table}

\begin{table}[!t]
    \centering
    \caption{Doubly-dispersive channel parameters.}
    \label{tab:channel_params}
    \begin{tabular}{@{}lcl@{}}
        \toprule
        \textbf{Parameter} & \textbf{Symbol} & \textbf{Value} \\
        \midrule
        Number of propagation paths
            & $R$ & $4$ \\
        Carrier frequency
            & $f_{\mathrm{c}}$ & $4~\mathrm{GHz}$ \\
        Maximum delay index
            & $\ell_{\max}$ & $3$ \\
        Maximum Doppler index
            & $f_{\max}$ & $2$ \\
             Carrier wavelength
            & $\lambda_c=c/f_c$ & $74.948~\mathrm{mm}$ \\
        Signal bandwidth
            & $B$ & $1~\mathrm{MHz}$ \\
        Sampling frequency
            & $f_s$ & $2~\mathrm{MHz}$ \\
        Sampling interval
            & $T_s$ & $0.5~\mu\mathrm{s}$ \\
        Maximum radial velocity
         & $v_{\max}$ & $2342.09 ~\mathrm{m/s}$ \\
       \bottomrule
    \end{tabular}
\end{table}

\subsection{Analysis of the \ac{AF} Statistics under \ac{HPA} Nonlinearities}
The statistical characterization developed in subsection~\ref{subsection:AF_under_PA_NL} provides a framework for interpreting the influence of \ac{HPA} nonlinearities on the \ac{AF}.
In the following, the mean and variance in \eqref{eq:mu_ycubic_extended} and \eqref{eq:var_y_cubic_extended} are analyzed by separating the contributions of the useful signal, governed by the Bussgang gain $\kappa$, and the nonlinear distortion, represented by the covariance matrix $\mathbf{R}_t$. The \ac{AF} sidelobe statistics are evaluated at a representative delay-Doppler bin, chosen as $(l_0,f_0)=(10,10)$.

\begin{figure*}[t]
  \centering
    \subfloat[Mean]{%
  \includegraphics[width=0.49\linewidth]{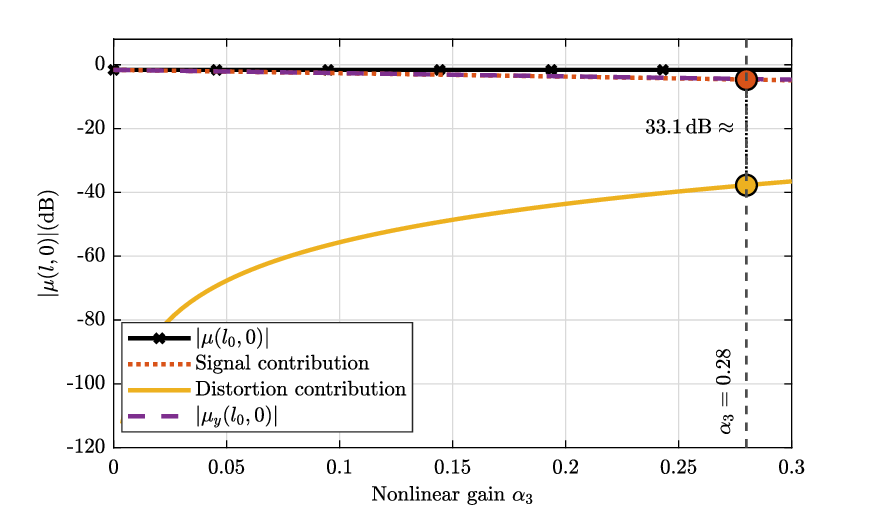}%
    }\hfil
    \subfloat[Variance]{%
  \includegraphics[width=0.49\linewidth]{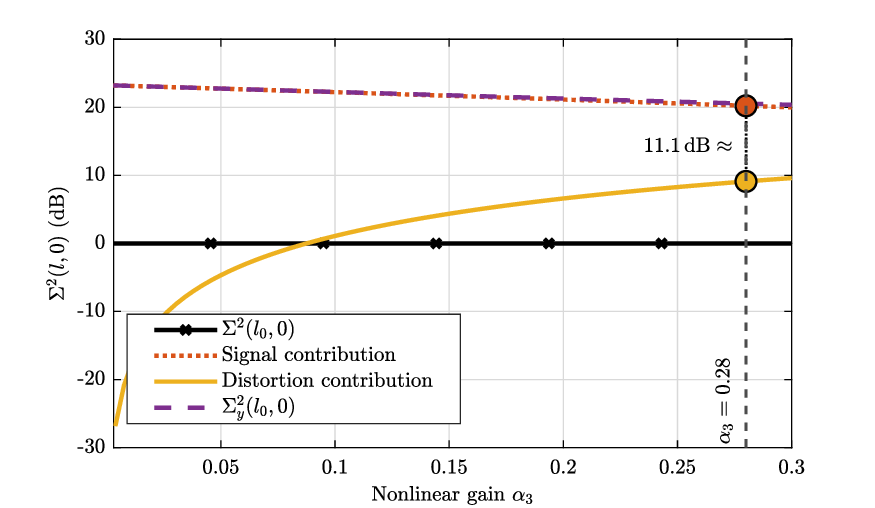}%
    }
    \caption{{Zero-Doppler (a) mean and (b) variance components, versus the nonlinear gain $\alpha_3$.}}
    \label{fig:zd_decomp}
\end{figure*}

Figs~\ref{fig:zd_decomp}.a and~\ref{fig:zd_decomp}.b show the decomposition of the \ac{AF} mean and variance for the zero-Doppler sidelobe normalized to the corresponding reference signal sidelobe value. The figures reveal that the delay sidelobe statistics remain largely governed by the useful signal component despite the presence of \ac{HPA} nonlinearities. As the nonlinearity increases, the useful contribution is attenuated through the Bussgang gain, whereas the distortion-induced contribution remains comparatively small. At the representative operating point, $e.g.~\alpha_3=0.28$ fitted to the considered Rapp AM--AM response corresponding to the \ac{IBO} = 4~dB operating point, the distortion-induced mean is more than 30~dB below the useful contribution, while the distortion variance is approximately 11.1~dB lower than the useful variance.

\begin{figure*}[t]
  \centering
    \subfloat[Mean]{%
  \includegraphics[width=0.49\linewidth]{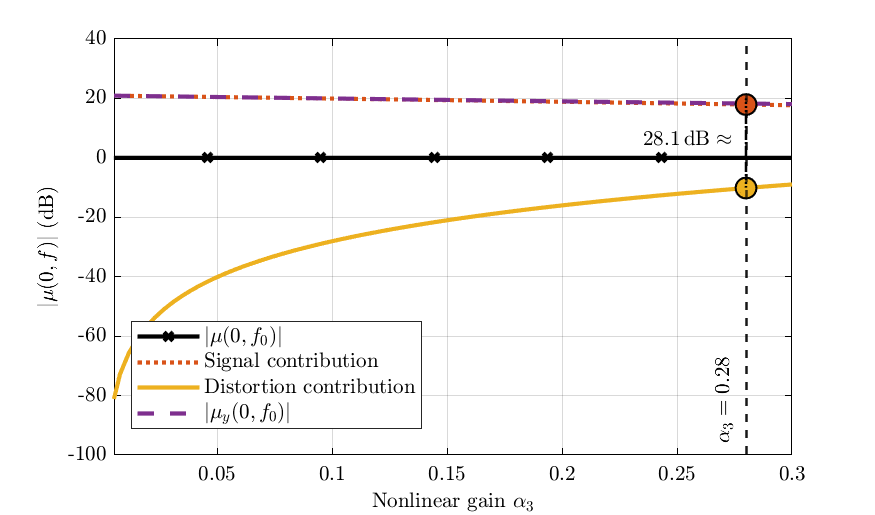}%
    }\hfil
    \subfloat[Variance]{%
  \includegraphics[width=0.49\linewidth]{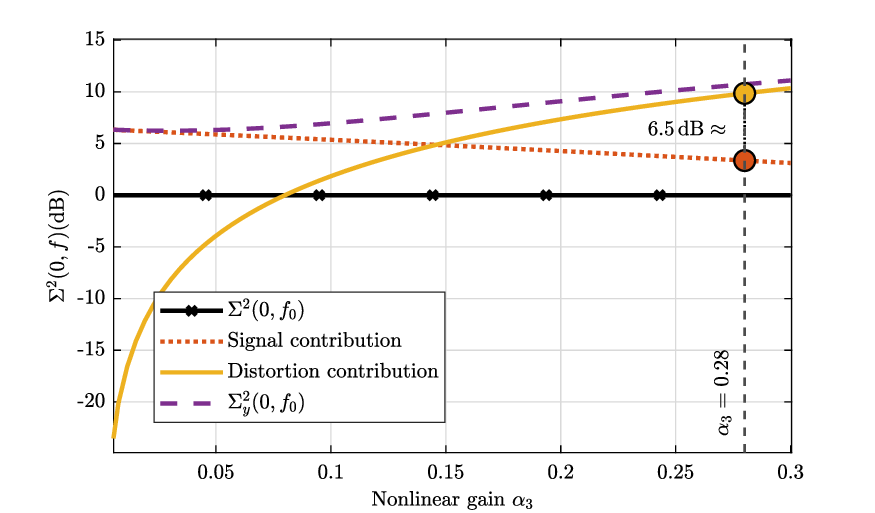}%
    }
    \caption{{Zero-delay (a) mean and (b) variance components, versus the nonlinear gain $\alpha_3$.}}
    \label{fig:zl_decomp}
\end{figure*}

Figs~\ref{fig:zl_decomp}.a and~\ref{fig:zl_decomp}.b show the decompositions of the zero-delay sidelobe normalized to the corresponding reference signal sidelobe value. The Doppler sidelobes exhibit a markedly different behavior. While the useful signal contribution is attenuated in the same manner through the Bussgang gain, the distortion-induced mean increases much more rapidly than in the zero-Doppler case, and the distortion exceeds the useful contribution as $\alpha_3$ increases. Therefore, \ac{HPA} nonlinearities affect the fluctuations of the Doppler sidelobes much more strongly than those of the delay sidelobes.

\begin{figure*}[t]
  \centering
    \subfloat[Mean]{%
  \includegraphics[width=0.49\linewidth]{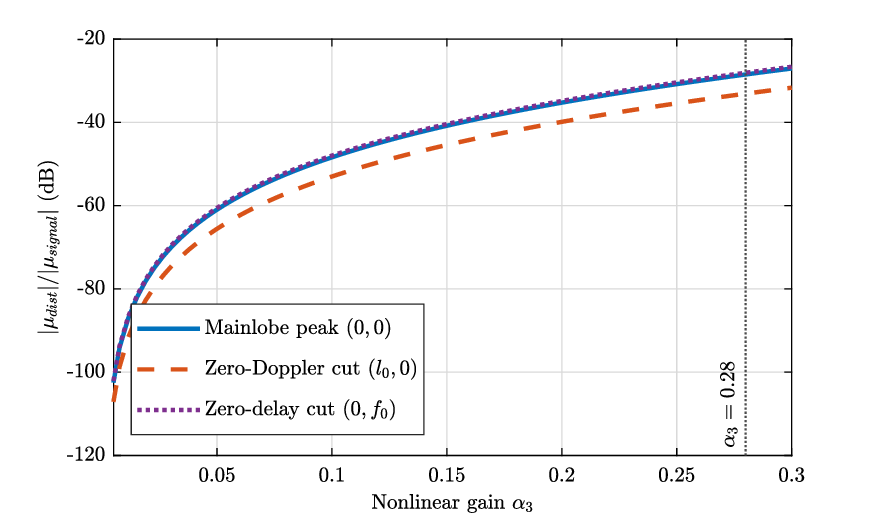}%
    }\hfil
    \subfloat[Variance]{%
  \includegraphics[width=0.49\linewidth]{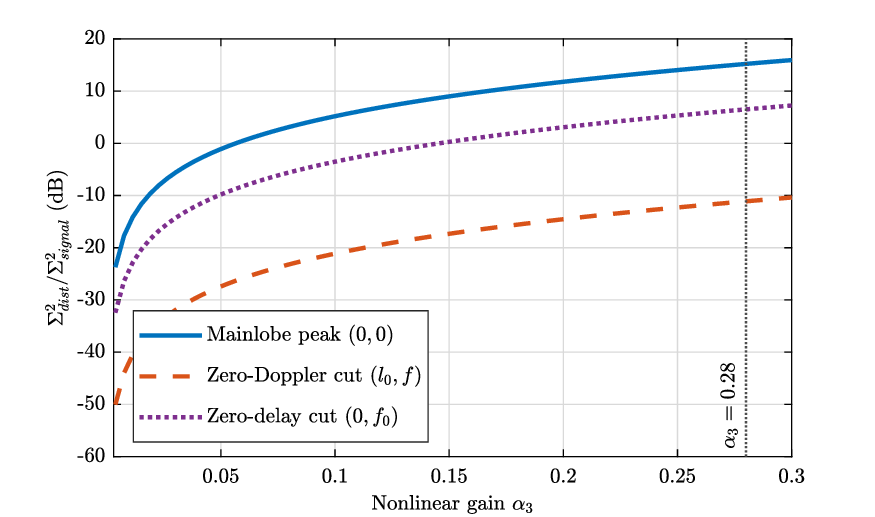}%
    }
    \caption{{Distortion-to-signal ratio of the mainlobe, zero-Doppler and zero-delay (a) mean and (b) variance, versus the nonlinear gain $\alpha_3$.}}
    \label{fig:ratio}
\end{figure*}
To compare the sensitivity of the \ac{AF} cuts independently of their absolute magnitudes, Figs~\ref{fig:ratio}.a and~\ref{fig:ratio}.b present the distortion-to-signal ratios for the mean and variance. These ratios quantify the relative contribution of the nonlinear distortion with respect to the useful signal component, enabling a direct comparison between the mainlobe peak and the delay- and Doppler-domain sidelobes. As the nonlinearity increases, the distortion-to-signal ratio rises for both the mean and the variance. Nevertheless, the zero-Doppler cut consistently exhibits the lowest ratios, confirming that the delay-domain sidelobes are the least sensitive to \ac{HPA} distortion. In contrast, the zero-delay cut curve is closer to the mainlobe curve, particularly in terms of the mean, indicating that the Doppler-domain sidelobes experience a distortion level closer to that of the mainlobe peak.

These ratios provide a statistical interpretation of the ambiguity metrics reported next. Let us recall that the \ac{PSLR} is defined as the ratio between the largest sidelobe magnitude and the mainlobe peak, whereas the \ac{ISLR} measures the ratio between the total sidelobe energy and the mainlobe energy \cite{rou2025normalized}. Consequently, both metrics depend on the relative evolution of the sidelobes with respect to the mainlobe. The previous ratio results are thus reflected by the \ac{AF}-related metrics presented in Table~\ref{tab:af_metrics_pa_NL}; the delay-domain \ac{PSLR} and \ac{ISLR} exhibit only marginal variations, while the largest changes are observed along the Doppler cut. Moreover, since the \ac{ISLR} is determined by the expected sidelobe energy through $\mathbb{E}[|\mathcal{A}(l,f)|^2]=|\mu(l,f)|^2+\Sigma^2(l,f)$, its evolution can be explained by the combined contributions of the mean and variance derived above. In particular, the decrease in Doppler-domain \ac{ISLR} reflects a nonlinear redistribution of the \ac{AF} energy that reduces the sidelobe-to-mainlobe energy ratio. These results confirm that AFBM preserves its superior localization in the Doppler domain under \ac{HPA} nonlinearities, whereas OFDM remains the waveform exhibiting the strongest localization in the delay domain. 
\begin{figure*}[t]
  \centering
    \subfloat[Zero-Doppler]{%
  \includegraphics[width=0.49\linewidth]{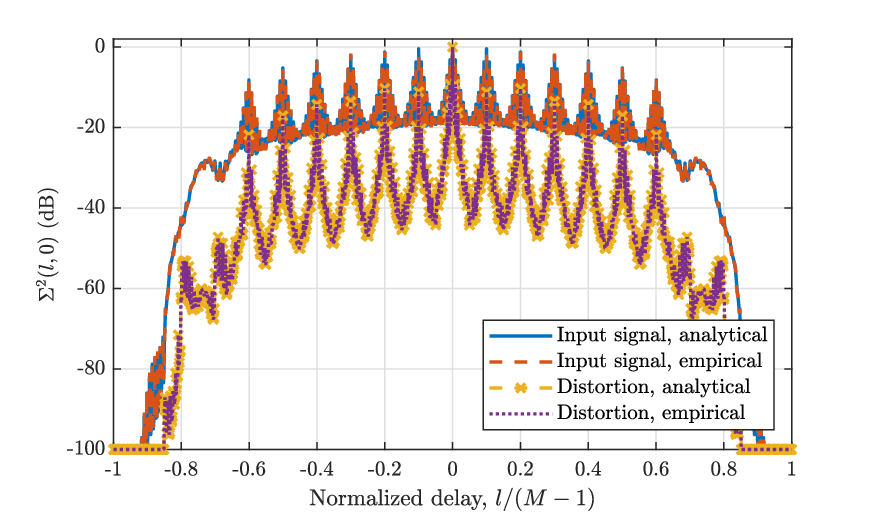}%
    }\hfil
    \subfloat[Zero-delay]{%
  \includegraphics[width=0.49\linewidth]{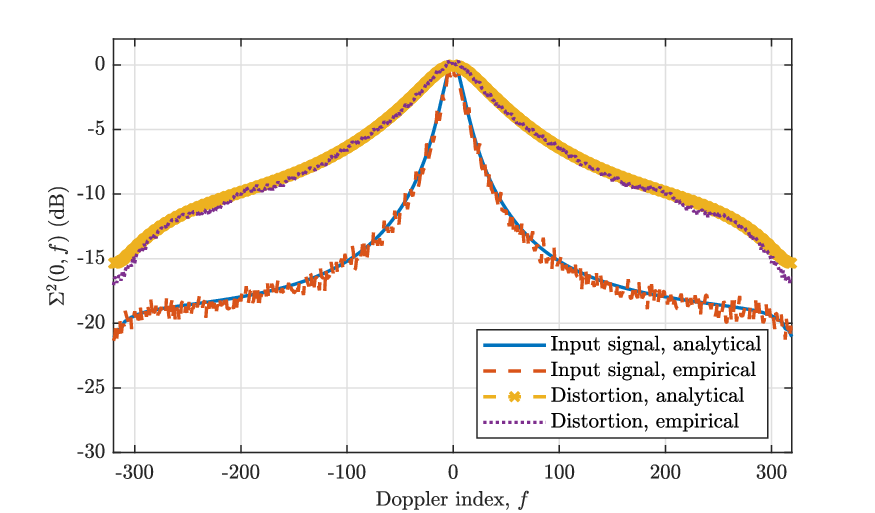}%
    }
    \caption{{Variance of the (a) zero-Doppler and (b) zero-delay cuts, under \ac{HPA} nonlinearities, \ac{IBO} = 4~dB.}}
    \label{fig:var_cuts}
\end{figure*}

Fig.~\ref{fig:var_cuts} shows the analytical zero-Doppler and zero-delay variance profiles for the nonlinear gain $\alpha_3 =0.28$. The delay-domain structure is largely preserved, whereas the Doppler-domain variance undergoes a more noticeable redistribution around the mainlobe, consistent with the pointwise analysis and simulations.

\begin{table*}[t]
\centering
\caption{PSLR and ISLR of the delay and Doppler \ac{AF} cuts before and after \ac{HPA} nonlinearities, \ac{IBO} = 4~dB.}
\label{tab:af_metrics_pa_NL}
\setlength{\tabcolsep}{12pt}
\begin{tabular}{llrrrr}
\toprule
\multirow{2}{*}{Waveform} 
& \multirow{2}{*}{Case} 
& \multicolumn{2}{c}{Delay cut} 
& \multicolumn{2}{c}{Doppler cut} \\
\cmidrule(lr){3-4}\cmidrule(lr){5-6}
& & PSLR (dB) & ISLR (dB) & PSLR (dB) & ISLR (dB) \\
\midrule
\multirow{2}{*}{AFBM}
& Input  & -18.76 &  0.03 & -13.48 & 3.48 \\
& Output & -18.81 &  0.14 & -13.44 & 1.94 \\
\addlinespace
\multirow{2}{*}{AFDM}
& Input  & -27.04 &  0.00 & -13.33 & 6.45 \\
& Output & -27.29 & -0.06 & -13.29 & 4.30 \\
\addlinespace
\multirow{2}{*}{OFDM}
& Input  & -33.05 & -4.75 & -13.14 & 6.35 \\
& Output & -33.26 & -4.62 & -13.34 & 4.32 \\
\bottomrule
\end{tabular}
\end{table*}

\subsection{Analysis of the BER under \ac{HPA} Nonlinearities}
The previous subsection demonstrated that the impact of \ac{HPA} nonlinearities on the \ac{AF} is governed by the interaction between the nonlinear distortion statistics and the AFBM modulation matrix. We now investigate whether the resulting robustness also extends to the communication performance.

To evaluate the impact of \ac{HPA} nonlinearities on communication performance, the AFBM receiver \ac{BER} was evaluated using the standard \ac{LMMSE} detector and the proposed GaBP and FOGa detectors, assuming that the receiver has perfect channel knowledge and knowledge of the Bussgang gain $\kappa$ and equivalent distortion variance $\sigma_d^2$. The Rapp model was considered in our simulations.

\begin{figure}[t]
  \centering
  \includegraphics[width=1\linewidth]{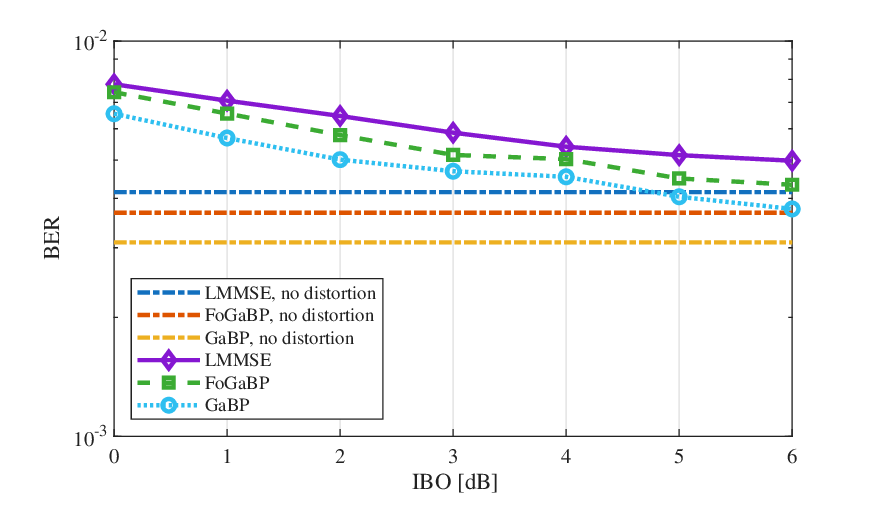}
  \vspace{-5ex}
     \caption{Comparison of the effect of varying \ac{IBO} values at a SNR of 15 dB for the considered detection schemes.}
    \label{fig:ber_vs_IBO}
\end{figure}

Figure~\ref{fig:ber_vs_IBO} shows the BER as a function of the \ac{IBO} at a fixed \ac{SNR} of $15$~dB and for a 4-QAM modulation. For all three detectors, the BER decreases as the \ac{IBO} increases due to the HPA operating farther from saturation. Using the no-distortion result of each detector as its respective reference, the BER increase at an \ac{IBO} of $1$~dB is approximately 41.31\% for LMMSE, 43.98\% for FOGa, and 45.56\% for the GaBP detector. All three detectors show similar degradation due to the HPA, but the GaBP detector outperforms the FOGa and \ac{LMMSE} in scenarios with and without HPA distortion.

\begin{figure*}[t]
  \centering
    \subfloat[\ac{IBO} = 1 dB]{%
  \includegraphics[width=0.49\linewidth]{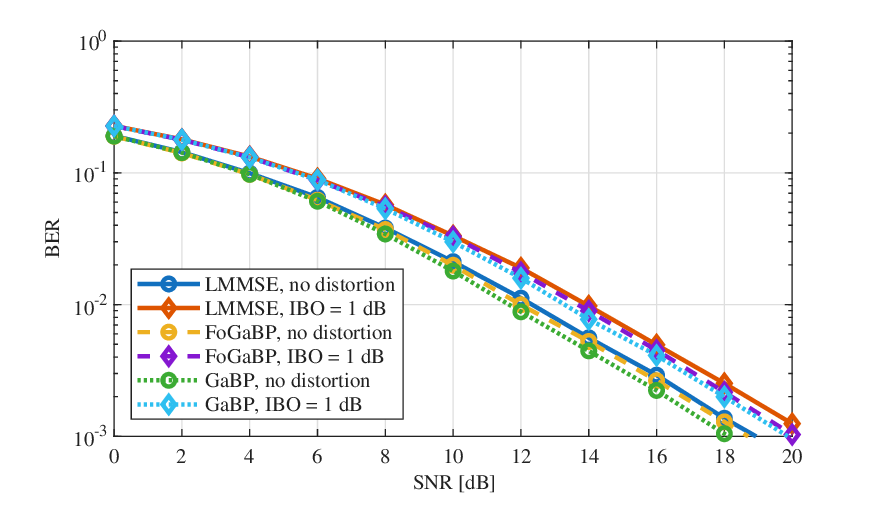}%
    }\hfil
    \subfloat[\ac{IBO} = 4 dB]{%
  \includegraphics[width=0.49\linewidth]{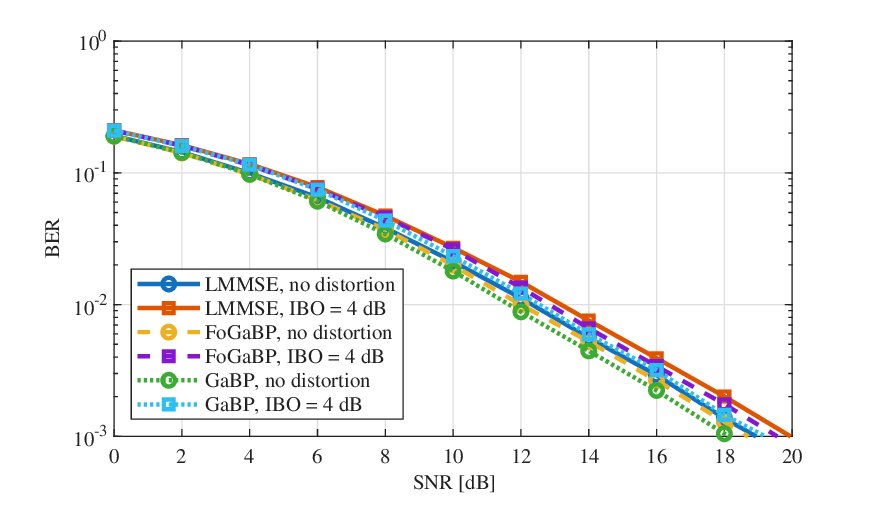}%
    }
    \caption{AFBM BER with 4-QAM on DD channel with the considered detection methods.}
    \label{fig:ber_dd_4}
\end{figure*}

\begin{figure*}[t]
  \centering
    \subfloat[\ac{IBO} = 1 dB]{%
  \includegraphics[width=0.49\linewidth]{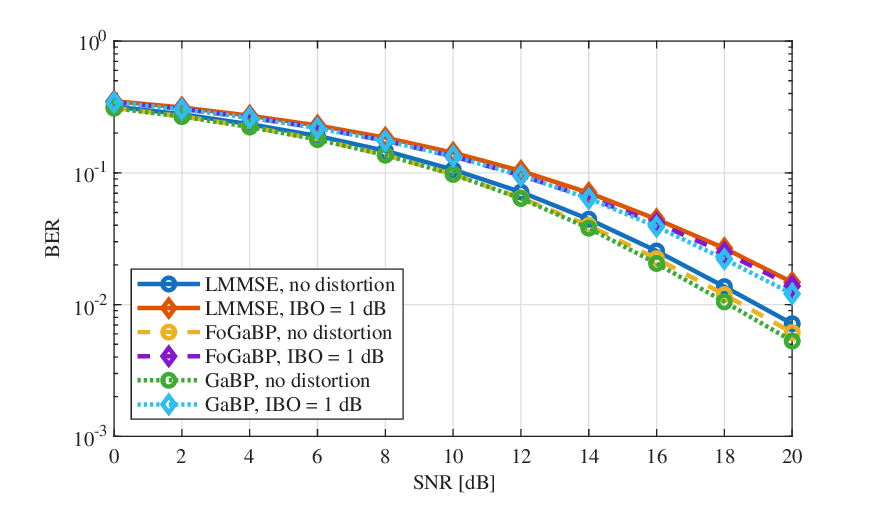}%
    }\hfil
    \subfloat[\ac{IBO} = 4 dB]{%
  \includegraphics[width=0.49\linewidth]{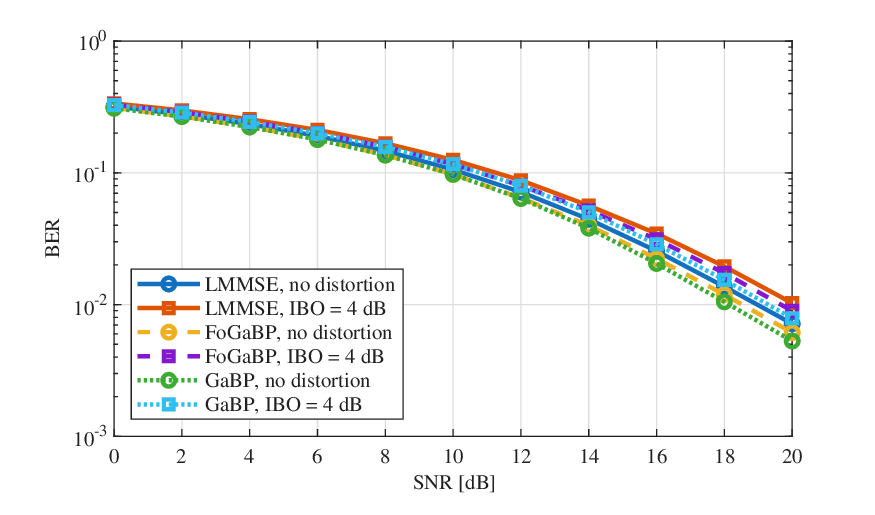}%
    }
     \caption{AFBM BER with 16-QAM on DD channel with the considered detection methods.}
    \label{fig:ber_dd_16}
\end{figure*}

Figs~\ref{fig:ber_dd_4} and~\ref{fig:ber_dd_16} show the BER as a function of \ac{SNR} for 4-QAM and 16-QAM, respectively. In the low values of SNR, the distorted and non-distorted curves remain close due to the performance being mostly limited by the doubly dispersive channel interference and thermal noise. At higher SNR values, the effect of the nonlinearity becomes more apparent due to the reduction in performance being mostly dominated by the HPA distortions. 
Regarding the choice of detection method, while FOGa exhibits slightly worse BER performance than the GaBP detector, it requires substantially fewer computations, corresponding to approximately $21\%$ and $7\%$ of GaBP complexity for 4-QAM and 16-QAM, respectively. This makes FOGa an attractive trade-off between detection performance and computational cost.

\section{Conclusion} \label{sec:conclusion}

This paper investigated the impact of \ac{HPA} nonlinearities on \ac{AFBM} from both sensing and communication perspectives. An approximate statistical characterization of the post-amplification \ac{AF} was derived by exploiting the Bussgang decomposition and a cubic approximation of the \ac{HPA}, yielding closed-form expressions for the mean and variance of the \ac{AF}. The proposed analysis showed that the effect of nonlinear distortion is governed by the interaction between the distortion statistics and the AFBM modulation matrix, explaining the different sensitivities of the delay- and Doppler-domain ambiguity characteristics.
The analytical results were validated through Monte Carlo simulations and shown to accurately predict the evolution of the \ac{AF} statistics as well as the resulting PSLR and ISLR. Next, a Gaussian belief propagation receiver accounting for the nonlinear distortion statistics was proposed. Numerical results over doubly dispersive channels showed that the proposed detector achieves a favorable complexity--performance trade-off while confirming that the inherent properties of AFBM enable robust communication performance under \ac{HPA} nonlinearities. In the follow-up, dedicated \ac{PAPR}-reduction techniques could further improve the hardware efficiency of \ac{AFBM} by enabling operation at lower \ac{IBO}. Moreover, the proposed distortion-aware communication receiver motivates the development of distortion-aware sensing receivers for enhanced sensing performance under \ac{HPA} nonlinearities.

\appendices
\section{Proof of Equations \eqref{eq:kappa_closed_form} and \eqref{eq:sigma_d_closed_form}} \label{proof:bussgang_poly}

To obtain closed-form expressions for the Bussgang parameters, the considered Rapp \ac{HPA} is approximated by an equivalent third-order polynomial over the input-amplitude range corresponding to the considered operating point. The equivalent polynomial is obtained by a least-squares fitting of its AM-AM characteristic to the Rapp response, yielding the model
\begin{equation*}
y(n)
=
a_1 s(n)
+
a_3 |s(n)|^2 s(n),
\label{eq:cubic_model}
\end{equation*}
where $
a_1=1$, $a_3=-\alpha_3$, and $\alpha_3>0$ denotes the nonlinear gain coefficient \cite{ismail2024robustness}. Assuming that the AFBM TD signal is circularly symmetric complex Gaussian \cite{gourar2026robustness}, its even-order moments satisfy $
\mathbb{E}
\!\left[
|s(n)|^{2m}
\right]
=
m!\sigma_s^{2m}$. Substituting these moments into the Bussgang decomposition yields the following,
\begin{equation*}
\kappa
=
1
-
2\alpha_3\sigma_s^2.
\end{equation*}

The corresponding distortion component is
\begin{align*}
d(n)
&=
y(n)-\kappa s(n) =
-\alpha_3
\left(
|s(n)|^2
-
2\sigma_s^2
\right)
s(n),
\end{align*}
with its variance expressed as
\begin{align*}
\sigma_d^2
&=
\mathbb{E}
\!\left[
|d(n)|^2
\right] =
\alpha_3^2
\mathbb{E}
\!\left[
|s(n)|^2
\left(
|s(n)|^2
-
2\sigma_s^2
\right)^2
\right]
=
2\alpha_3^2
\sigma_s^6.
\end{align*}

\bibliographystyle{IEEEtran}
\bibliography{biblio.bib}

\end{document}